

\newif\ifnomssymb \nomssymbtrue 
\newread\amsfile
\immediate\openin\amsfile=mssymb.tex
\immediate\ifeof\amsfile
 \else\closein\amsfile\input mssymb.tex\nomssymbfalse\fi
\closein\amsfile


\tolerance=3000 \hbadness=3000 \hfuzz=6pt
\magnification=1200
\baselineskip=16pt plus 2pt minus 1pt

\font\sevteenrm=cmr17
\font\twelverm=cmr12
\font\eightrm=cmr8
\def\rm{\fam0 \tenrm}
\overfullrule=0pt



\newcount\yearltd\yearltd=\year\advance\yearltd by -1900
{\count255=\time\divide\count255 by 60 \xdef\hourmin{\number\count255}
  \multiply\count255 by-60\advance\count255 by\time
  \xdef\hourmin{\hourmin:\ifnum\count255<10 0\fi\the\count255}}

\newwrite\ffile  
\newif\iffclosed \fclosedtrue 
\def\fwrite#1{\iffclosed\immediate\openout\ffile=\jobname.fwd\fclosedfalse\fi
              \immediate\write\ffile{#1}}


\def\freference#1{\fwrite{\noexpand\xdef\noexpand#1\noexpand{#1}}}

\newread\tfile
\def\testinput#1{
                 \immediate\openin\tfile=#1
                 \immediate\ifeof\tfile\else\closein\tfile\input#1\fi
                 \closein\tfile
                  }
\xdef \Sintro {{1}}
\xdef \Sprop {{2}}
\xdef \Spert {{3}}
\xdef \SLT {{4}}
\xdef \SSPLT {{4.1}}
\xdef \singlt {{\hbox {$(4.27)$}}}
\xdef \ntm {{\hbox {$(4.28)$}}}
\xdef \SSDLT {{4.2}}
\xdef \SSSLT {{4.3}}
\xdef \SBV {{5}}
\xdef \SSDMV {{5.1}}
\xdef \SSPMV {{5.2}}
\xdef \SSSMV {{5.3}}
\xdef \SSCMV {{5.4}}
\xdef \Stheorem {{6}}
\xdef \Appendix {{Appendix}}

\testinput{\jobname.fwd}



%


\def\eqlabform#1{{\hbox{$(\chapsyme\secsyme#1)$}}}

\def\thmlabform#1{{$\chapsyme\secsyme#1$}}
\def\figlabform#1{{$#1$}}

\def\nolabels{\def\eqnlabel##1{}\def\eqlabel##1{}\def\reflabel##1{}}
\def\writelabels{\def\eqnlabel##1{%
{\escapechar=` \hfill\rlap{\hskip.09in\string##1}}}%
\def\eqlabel##1{{\escapechar=` \rlap{\hskip.09in\string##1}}}%
\def\reflabel##1{\noexpand\llap{\string\string\string##1\hskip.31in}}}
\nolabels

\def\eqname#1{\global\advance\meqno by1
               \xdef #1{\eqlabform{\the\meqno}} }
\def\eqaname#1{\global\advance\meqno by1
               \xdef #1##1{{\eqlabform{\the\meqno##1}}} }

\def\eqna#1{\eqaname#1\eqnlabel{#1$\{\}$}}

\def\etag#1{\eqname#1\eqno#1\eqnlabel#1}

\def\eqnt#1{\global\advance\meqno by1 \xdef #1##1{\hbox{$(T\the\meqno##1)$}}%
            \eqnlabel{#1$\{\}$}}

\def\ftag#1{\etag{#1}\freference{#1}}

\def\newthm#1{\global\advance\thmno by
1\xdef#1{{\thmlabform{\the\thmno}}}#1\freference{#1}}
\def\newfig#1#2#3{
\global\advance\figno by 1\xdef#1{{\figlabform{\the\figno}}}\freference{#1}
\bigskip
\psfig{file=#2}
\nobreak\medskip
\noindent{\bf Figure #1.  }{\it  #3}
\medskip
}

\catcode `@=11
\newtoks\@temptokena  

\newif\if@afterindent \@afterindenttrue
\newdimen\@tempdima
\def\contentdepth{1}
\def\@pnumwidth{20pt}
\def\@tocrmarg{80pt}
\def\@dotsep{1.7}

%
%
\def\dottedtocline#1#2#3#4#5{\ifnum #1>\contentdepth \else
  \vskip \z@ plus .2pt
  {\leftskip #2\relax \rightskip \@tocrmarg \parfillskip -\rightskip
    \parindent #2\relax\@afterindenttrue
   \interlinepenalty\@M
   \leavevmode
   \@tempdima #3\relax \advance\leftskip \@tempdima \hbox{}\hskip -\leftskip
   #4\nobreak\leaders\hbox{$\m@th \mkern \@dotsep mu.\mkern \@dotsep
       mu$}\hfill \nobreak \hbox to\@pnumwidth{\hfil\rm #5}\par}\fi}
\def\numberline#1{\hbox to\@tempdima{#1\hfil}}

\xdef\chapsym{}\xdef\chapsyme{} 
\xdef\secsym{}\xdef\secsyme{}
\global\newcount\chapno \global\chapno=0
\global\newcount\secno \global\secno=0
\global\newcount\subsecno \global\subsecno=0
\global\newcount\meqno \global\meqno=0
\global\newcount\thmno \global\thmno=0
\global\newcount\figno \global\figno=0
\global\newcount\ftno \global\ftno=0
\def\ourfolio{\ifnum\pageno<0 \romannumeral-\pageno \else\number\pageno\fi}
\def\leaderfill{\leaders\hbox to 1em{\hss . \hss}\hfill}

\newwrite\cfile  
\newif\ifcclosed \cclosedtrue 
\def\cwrite#1{\ifcclosed\immediate\openout\cfile=\jobname.con\cclosedfalse\fi
              \immediate\write\cfile{#1}}

\newwrite\pfile  
\newif\ifpclosed \pclosedtrue
\newcount\pno   \pno=0  
\def\pwrite#1{\ifpclosed\immediate\openout\pfile=\jobname.pre\pclosedfalse\fi
              \immediate\write\pfile{#1}}

\def\extrahead#1{\vfil\eject\centerline{#1}\bigskip}

\def\extra#1{\extrahead{#1} \@temptokena{#1}
  \cwrite{\noexpand \dottedtocline{0}{0pt}{20pt}
  {\the\@temptokena}{\ourfolio} } }
\def\extraearly#1{\extrahead{#1}\@temptokena{#1}
  \pwrite{\noexpand \dottedtocline{0}{0pt}{20pt}
  {\the\@temptokena}{\ourfolio}} }

\def\chapter#1#2#3{
\global\ftno=1
\xdef#1{#2}
\xdef\chapsym{#2}
\xdef\chapsyme{#2.}
\global\thmno=0\global\figno=0\global\meqno=0\global\secno=0\global\subsecno=0
\vfill\eject
\noindent{\bf \chapsyme\ #3}
\@temptokena{#3}
\cwrite{\noexpand
        \dottedtocline{0}{0pt}{15pt}
           {\noexpand\numberline{#2.}{\the\@temptokena}}{\ourfolio}}
\freference{#1}
\par\nobreak\medskip\nobreak
}

\def\newchap#1#2{\global\advance\chapno by 1
                 \chapter#1{\the\chapno}{{#2}}
                 }

\def\section#1#2#3{
\xdef\secsym{\chapsyme#2}
\xdef\secsyme{\secsym.}
\xdef#1{{\secsym}}
\global\subsecno=0
\bigbreak\medskip
\@temptokena{#3}
\noindent{\bf\secsyme\ #3 }
\cwrite{\noexpand
        \dottedtocline{1}{15pt}{25pt}
              {\noexpand\numberline{\secsyme}{\the\@temptokena}}{\ourfolio}}
\freference{#1} }
\def\newsec#1#2{\global\advance\secno by 1
                 \section#1{\the\secno}{{#2}}
                 }

\def\subsection#1#2#3{
\xdef#1{{\secsyme#2}}
\medbreak\smallskip
\noindent{\bf\secsyme#2 #3 }
\@temptokena{#3}
\cwrite{\noexpand
        \dottedtocline{2}{40pt}{25pt}
           {\noexpand\numberline{\secsyme#2.}{\the\@temptokena}}{\ourfolio}}
\freference{#1}
}
\def\newsubsec#1#2{\global\advance\subsecno by 1
                   \subsection#1{\the\subsecno}{{#2}}}

\def\subsubsec#1{{\medbreak\smallskip\noindent{\bf{#1:}}}}

\def\listtoc{\vfill\eject
             \immediate\closeout\pfile
             \centerline{{\bf Table of Contents}}\bigskip
             \testinput{\jobname.pre}
             \testinput{\jobname.con}
             \vfill\eject
              }

\def\footfont{\eightrm}
\newskip\footskip\footskip12pt plus 1pt minus 1pt 
\def\f@@t{\footfont\baselineskip\footskip\bgroup\aftergroup\@foot\let\next}
\setbox\strutbox=\hbox{\vrule height9.5pt depth4.5pt width0pt}
\def\foot{\global\advance\ftno by1\footnote{$^{\the\ftno}$}}
%
\newwrite\ftfile
\def\footend{\def\foot{\global\advance\ftno by1\chardef\wfile=\ftfile
$^{\the\ftno}$\ifnum\ftno=1\immediate\openout\ftfile=foots.tmp\fi%
\immediate\write\ftfile{\noexpand\smallskip%
\noexpand\item{f\the\ftno:\ }\pctsign}\findarg}%
\def\footatend{\vfill\eject\immediate\closeout\ftfile{\parindent=20pt
\centerline{\bf Footnotes}\nobreak\bigskip\input foots.tmp }}}
\def\footatend{}

\catcode `@=12


\global\newcount\refno \global\refno=1
\newwrite\rfile

\def\ref#1#2{
    \nref#1{#2}}
\def\nref#1#2{\xdef#1{{\the\refno}}%
\ifnum\refno=1\immediate\openout\rfile=\jobname.ref\fi%
\immediate\write\rfile{\noexpand\item{[#1]\ }\reflabel{#1}#2}%
\global\advance\refno by1}
\def\addref#1{\immediate\write\rfile{\noexpand\item{}#1}}

\def\semi{;\hfil\noexpand\break}

\def\listrefs{\immediate\closeout\rfile
              \parindent=20pt\baselineskip=14pt
              \extra{References}
              {\frenchspacing%
               \testinput{\jobname.ref}\vfill\eject}
              \nonfrenchspacing
              }


\def\book#1#2#3{{\it #1}, #2, #3}

\def\cbook#1#2#3#4{{\it #1}, #2, #3, #4}

\def\article#1#2#3#4#5{{\it #1}, #2 {\bf #3} (#4) #5}

\def\preprint#1#2{{\it #1}, #2}




\ifnomssymb
\def\IR{{\hbox{{\rm I}\kern-.2em\hbox{\rm R}}}}
\def\IB{{\hbox{{\rm I}\kern-.2em\hbox{\rm B}}}}
\def\IC{{\ \hbox{{\rm I}\kern-.6em\hbox{& ams\bf C}}}}
\def\IQ{{\ \hbox{{\rm I}\kern-.54em\hbox{\bf Q}}}}
\def\IZ{{\hbox{{\rm Z}\kern-.4em\hbox{\rm Z}}}}
\def\id{{\hbox{{\rm 1}\kern-.35em\hbox{\rm 1}} }}
\def\subsetneqq{{
\ \lower-2.2pt\vbox{\hbox{\rlap{$\subset$}\lower7.2pt\vbox{\hbox{$\neq$}}}}\ }}
\else
\def\IR{{\Bbb R}} \def\IC{{\Bbb C}} \def\IQ{{\Bbb Q}} \def\IZ{{\Bbb Z}}
\def\id{{\Bbb 1}}
\fi

\def\Cal{\cal}
\def\bold#1{{{{\bf #1}}}}
\def\cal{\fam2 } \def\rm{\fam0 \tenrm} 
\def\u#1{{\underline{#1}}}

\def\Bf{\bold}
\def\tl#1{\tilde#1}
\def\Underline#1{{$\underline{\hbox{#1}}$}}

\def\bnoalign#1{\noalign{\baselineskip=0pt#1}}
\def\casespace{\noalign{\vskip6pt}} 


 \def\cB{{\Cal{B}}} \def\cC{{\Cal{C}}} 
   
\def\cI{{\Cal{I}}}   \def\cL{{\Cal{L}}}
  \def\cO{{\Cal{O}}} 
  \def\cS{{\Cal{S}}} \def\cT{{\Cal{T}}}

\def\id{{ \hbox{{\rm 1}\kern-.35em\hbox{\rm 1}} }}
\def\Sum{\sum}   
\def\union{\bigcup}
\def\intersect{\bigcap}
\def\half{{ 1 \over 2}}
\def\quarter{{ 1 \over 4}}
\def\lap{\triangle}
\def\del{\partial}

\def\part#1{{\partial\over\partial #1}}
\def\partt#1{{\textstyle\part{#1}}}  
\def\tot#1{{d \over d #1}}

\def\dirac{{\hbox{$\partial$\kern-.53em$/$}}}

\def\dual{{^{\vee}}}
\def\endproof{$\rbrack\!\rbrack$}

\catcode`\@=11
\def\FN@{\futurelet\next}
\def\DN@{\def\next@}
\def\relaxnext@{\let\next\relax}
\def\nolimits@{\relaxnext@
 \DN@{\ifx\next\limits\DN@\limits{\nolimits}\else
  \let\next@\nolimits\fi\next@}%
 \FN@\next@}
\def\newmcodes@{\mathcode`\'"27\mathcode`\*"2A\mathcode`\."613A%
 \mathcode`\-"2D\mathcode`\/"2F\mathcode`\:"603A }
\def\operatorname#1{\mathop{\newmcodes@\kern\z@\fam\z@#1}\nolimits@}
\catcode`\@=\active
\def\op{\operatorname}

\def\Im{{  \op{ Im } }}  \def\Hom{{  \op{Hom} }}
\def\Lie{{ \op{ Lie} }}  
  \def\Tr{{  \op{Tr  } }}
\def\Map{{ \op{Map } }}  
  
 \def\ln{{  \op{ln  } }} \def\Met{{ \op{Met}  }}
\def\Sym{{ \op{ Sym} }} \def\ord{{\op{ord}}}



\ref\AS{S.\ Axelrod and I.\ M.\ Singer,
{\it Chern--Simons Perturbation Theory}, Proc. XXth DGM Conference
(New York, 1991) (S. Catto and A. Rocha, eds) World Scientific,
1992, 3--45.}
\ref\AXE{S.\ Axelrod, in preparation.}
\ref\BG{
A. Beilinson and V. Ginzburg,
  \article{Infinitesimal Structure of Moduli Spaces of $G$--bundles}
   {Intern. Mathem. Research Notices, Duke Math. J.}{4}{1992}{63-64}.
A. Beilinson, V. Drinfeld, and V. Ginzburg,
  \preprint{Differential Operators on Moduli Spaces of $G$--Bundles}
  {Preprint (1991)}.}
\ref\BGV{N. Berline, E. Getzler and M. Vergne,
\cbook{Heat kernels and Dirac operators}
{Springer Verlag}{Berlin-Heidelberg-New York}{1992}.}
\ref\FM{W.\ Fulton and R.\ MacPherson,
   \preprint{A Compactification of Configuration Space}
    {To appear in Annals of Math}.}
\ref\GET{E. Getzler,
   \article{A Short Proof of the Local Atiyah--Singer Index Theoerem}
{Topology}{25}{1986}{11-117}.}
\ref\Ho{L.\ Hormander,
 \book{Linear partial differential operators {\bf 3 }}
   {Springer--Verlag}{1985}.}
\ref\KON{M. Kontsevich, to appear.}
\ref\MATHAI{V. Mathai,
  \article{Heat Kernels and Thom Forms}{J. Func. Anal.}{104}{1992}{34-46}.}
\ref\WiI{E.\ Witten,
 \article{Quantum field theory and the Jones polynomial}
   {Comm.  Math. Phys.}{121}{1989}{351-399}.}

\ref\WiIV{E.\ Witten,
   \preprint{Chern--Simons Gauge Theory as a String Theory}
     {IAS preprint IASSNS-HEP-92/45 (1992)}.}

\pageno=1

{\hfill hep-th/9304087}
\vskip 1in
{\sevteenrm \centerline{Chern--Simons Perturbation Theory II} }
\vskip .75in
{\twelverm
\centerline{Scott Axelrod}\par
\par
\centerline{I.M.~Singer}\par
\par
\centerline{\it Math Department, MIT}\par
 }
\vskip 1.5in
{\hsize= 5.5 true in \hoffset= .5 true in
\noindent{\bf Abstract.}
In a previous paper [\AS], we used superspace techniques
to prove that perturbation theory
(around a classical solution with no zero modes) for
Chern--Simons quantum field theory on a general $3$-manifold $M$ is finite.
We conjectured (and proved for the case of $2$-loops) that, after adding
counterterms of the expected form, the terms in the perturbation theory
define topological invariants.  In this paper we prove this conjecture.
Our proof uses a geometric compactification of the region
on which the Feynman integrand of Feynman diagrams is smooth as well
as an extension of the basic propagator of the theory.
}
\vfill
\footnote{}{
This work was supported in part by the Divisions of Applied
Mathematics of the U.~S.~Department of Energy under contracts
DE-FG02-88ER25065 and 
DE-FG02-88ER25066.  
}
\vfil\eject
\def\Dz{D}			
\def\dlz{D^{\dagger}{}}		
\def\adp{{\bf g}}		

\def\cs{{CS}}			
\def\G{{\Bf{G}}}		
\def\LG{{\underline{\G}}}	
\def\uV{{\u{V}}}
\def\Bl{{ \op{ Bl} }}
\def\TM{{\widetilde{TM} }}
\def\lapp{{\lap_M}}
\def\BOmega{{\Omega}}
\def\flap{{\cO}}
\def\tflap{{\tl\flap}}  
\def\lmg{\Lambda^*_g}


\newsec\Sintro{Introduction}
\medskip

In a previous paper [\AS], we considered the perturbative expansion
for three dimensional Chern-Simons quantum field theory about a solution $A_0$
to the equations of motion.  We defined what we meant by the perturbative
expansion and showed perturbation theory was finite.
We showed that the first term in the pertubative expansion beyond the
semiclassical limit defines a geometric
invariant precisely in the manner one would expect
based on Witten's exact solution [\WiI].
We conjectured and gave strong evidence that the higher terms in the
expansion were geometric invariants of the same type.
In this paper we prove this conjecture.

More specifically, we take $A_0$ to be a flat connection on a principal
bundle $P$ with a compact structure group $G$ and a closed, oriented,
$3$-dimensional base $M$.  We also assume that $A_0$ has no zero modes,
i.e. that the cohomology of the exterior derivative operator
$\Dz:\Omega^*(M,\adp)\rightarrow \Omega^{*+1}(M,\adp)$, coupled to
the adjoint bundle $\adp$ of $P$ and $A_0$, vanishes.
By rewriting the Lorentz gauge fixed theory as a superspace theory in [\AS],
we
were able to obtain Feynman rules that could be translated succinctly
into the language of differential forms.
To define the gauge fixing it was necessary to choose a
Riemannian metric $g$ on $M$.
For $l\ge 2$, the $l^{th}$ order term $I_l(M,A_0,g)$ in the pertubative
expansion is a multiple integral over
$M^{V}$, with $V=2(l-1)$, of a top form depending on $g$.
This top form,  the ``Feynman integrand'', is smooth on the open submanifold
$M^V_0\subset M^V$ consisting of the points away from all
diagonals, but is singular near the diagonals.
It is constructed
from products of the basic ``propagator'' $L$, the integral kernel
for the ``Hodge theory inverse'' to $\Dz$.
We showed that, despite
the singularities, the integral defining $I_l(M,A_0,g)$ is finite.
Also, we gave a ``formal proof of metric independence'' of $I_l(M,A_0,g)$
(ignoring the problem of products of singularities).   The only dependence
on the metric is therefore due to quantum field theoretic ``anomalies'',
which arise because of the behavior of the integrand near
$M^V\setminus M^V_0$.

The quantity $I_l$ decomposes as a sum of ``Feynman amplitudes'' for
trivalent graphs with $V$ vertices.  The nature of the anomalies is
most simply stated in terms of the piece $I_l^{conn}$ of $I_l$ which
comes from the sum over connected graphs.
We conjectured, and proved for $l=2$, that the dependence on the metric could
be cancelled by subtracting a multiple of the Chern-Simons invariant for the
metric connection.  This conjecture is proved for all $l$
in the present paper.

We analyzed the variation of $I_2$ with respect to a metric in
[\AS] by using Stokes theorem on the differential geometric blowup of
$M^2\setminus\Delta$ along the diagonal $\Delta$.  That space
$\Bl(M^2,\Delta)$ (see \S\Sprop) has a boundary which can be identified
with the tangent sphere bundle over $M$.  To extend the argument and prove
the theorem we will use a ``geometric blowup'' of $M^V$ along
$M^V\setminus M^V_0$.  This blowup $M[V]$ is a manifold with corners
and is a compactification of $M^V_0$
to which the Feynman integrand extends smoothly.
Our results can also be proved without
introducing $M[V]$ by using power counting arguments of the form found
in [\AS], but the use of $M[V]$ is more geometrical.
As we will explain below, $M[V]$ is the differential geometric analog of the
algebraic geometric compactification defined in [\FM] and [\BG].
Other compactifications besides $M[V]$ may also be employed to the same
end, but it would take us too far afield to explain this here.
In a private discussion, M. Kontsevich explained his use of $M[V]$ in
his work on Chern-Simons perturbation theory [\KON].  The appearance of
[\FM] and [\BG] convinced us that this approach would be the simplest.

We will also introduce an ``extended propagator''
$\tilde L$, a vector bundle valued form on $(M^2\setminus\Delta)\times\Met$,
where $\Met$ is the space of Riemannian metrics on $M$.
Readers worried about infinite dimensional spaces may take
$\Met$ to be any finite dimensional submanifold of the space of metrics.
Actually, for the proof of our main theorem,
we could equally well proceed by taking $\Met$ to be an interval
in the space of metrics.
However $\tilde L$ allows, among other things, an extension of the theory
to families of manifolds of any dimension, as will be shown in [\AXE].
This extension gives a mathematically precise version of
the ``field theory limit'' of the topological open string model considered
in [\WiIV].  It is also closely related to
ideas of M. Kontsevich [\KON].

$\tilde L$ may be expanded as a sum of its
pieces $\tilde L^{(d)}$ of homogeneous degree $d$ on $\Met$,
$$\tilde L = \tilde L^{(0)}+\tilde L^{(1)}+\tilde L^{(2)}
.\etag\last$$
The piece $\tilde L^{(0)}$ is just the original propagator $L$, considered
as a $2$-form on $M^2\times\Met$ of degree $0$ (i.e. an ordinary
function) in the $\Met$ directions.

As with $M[V]$, our  introduction of $\tilde L$ is also not strictly
necessary. One could express our discussion entirely in terms of
the separate components $\tilde L^{(0)}$ and $\tilde L^{(1)}$ of $\Met$,
without unifying them as part of a larger structure.
Although introducing $\tilde L$ will allow us to be more succinct, the
reader may find it illuminating to make the occurrences of
$L=\tilde L^{(0)}$ and $\tilde L^{(1)}$ explicit.  This will give the
arguments more in the language of [\AS], where $\tilde L^{(1)}$ is called
$B$.

\medbreak\bigskip

\noindent{\bf Outline:}

Sections \Sprop\ and \Spert\ are largely an exposition of parts of
[\AS] with some extensions and modifications, along with special accommodation,
we hope, to mathematicians.  See [\AS] and references therein
for more explanation of the relation to the physics literature.
We review the basic propagator $L$
and its properties in \S\Sprop.  In \S\Spert\ we define the terms in the
perturbation expansion, namely $I_l$ and $I_l^{conn}$, and give the
Feynman graph interpretation of these multiple integrals over $M^V$.

The properties of the extended propagator $\tilde L$ needed in the proof
of our main theorem are stated in \S\SSPLT.
The actual definition of $\tilde L$ and the proof of
some of the properties are given in \S\SSDLT.
The remaining properties, relating to the fact that $\tilde L$ extends
smoothly to a covariantly closed form on $\Bl(M^2,\Delta)\times\Met$,
are proved in \S\SSSLT.

The compactification $M[V]$ is described as a closure of $M^V_0$ in a larger
topological space in \S\SSDMV.
$M[V]$ is described explicitly as a point set in \S\SSPMV.
A stratification of $M[V]$ is introduced in \S\SSSMV.
One proof that $M[V]$ is a manifold with corners (such that the codimension $k$
open strata of the stratification are smooth open subsets of the
codimension $k$ boundary of $M[V]$) follows by directly mimicking
the construction in [\FM] but using differential geometric blowups rather
than algebraic geometric ones.
As an alternative to this, we give an explicit atlas of coordinates
on $M[V]$ in \S\SSCMV.

The results of
\S\SLT\ and \S\SBV\ allow us to prove the main theorem in \S\Stheorem.

A short appendix is included to describe our use of graded tensor
product and our mathematically unusual sign conventions for push-forward
integrals (which arise naturally from the superspace formulation of the field
theory).

The presentations in \S\SSSLT\ and \S\SSCMV\ are rather brief.
Further elaboration, in the context of generalizations,
will be found in a future paper by the first named author [\AXE].

\bigskip
\newsec\Sprop{Review of the Basic Propagator and It's Properties}
\medskip

The Feynman rules expressed in the language of differential forms use
the ``Hodge theory inverse'' to $\Dz$.  This is the operator
$$\Dz^{-1}\equiv
  \dlz\circ \lapp^{-1}=
 \lapp^{-1}\circ\dlz:\Omega^j(M,\adp)\rightarrow \Omega^{j-1}(M,\adp),
\ j=1,2,3
.\etag\last$$
Here $\dlz$ is the adjoint of $\Dz$ and $\lapp\equiv \{\Dz,\dlz\}$
is the associated Laplacian operator.  Adjoints are defined
with respect to the inner product on $\Omega^*(M,\adp)$ induced
by a choice of bi-invariant inner
product $<\,,\,>_{\Lie(G)}$ on the Lie algebra $\Lie(G)$ of $G$, and a choice
of Riemannian metric $g$ on $M$.

The operator $\Dz^{-1}$ can be written as an integral operator with
kernel $L$, known as the propagator.  $L$
belongs to $\Omega^2(M_1\times M_2,\adp_1\otimes\adp_2)$
(where the subscripts $1$ and $2$ refer to distinct copies of $M$ and
the corresponding bundles over them), and is defined by
$$
(\Dz^{-1}\psi)_a(x) = \int_{y\in M_2} L_{ab}(x,y)\wedge\psi_b(y)
\quad\forall\psi\in\Omega^*(M,\adp)
.\etag\defL$$
Here we have introduced the Lie algebra indices $a$ and $b$
which arise after introducing an orthonormal basis $\{T_a\}$ for $\Lie(G)$
and a local trivialization of $P$.\foot{
   Note that we have not used the more usual pairing
$\int_{y\in M_2} L_{ab}(x,y)\wedge *\psi_b(y)$.  Using the metric on
$\Lie(G)$ to identify $\adp_1\otimes\adp_2$ with $\Hom(\adp_2,\adp_1)$,
$L(x,y)\wedge\psi(y)$ means wedge the forms and apply the linear
transformation from $\adp_2$ to $\adp_1$.
}
The totally anti--symmetric structure constants $f_{abc}$ for $G$
are given by $[T_a,T_b] = f_{abc} T_c$.

The relation between operators and their associated integral kernels
used in \defL\ is the one that arises naturally from the superspace
formalism.  This gives an unusual sign convention in pushforward
integrals like the one in \defL.
Using these sign conventions (see the appendix for more details),
the relation
$$ \int \psi_a\wedge (D \phi)_a
= (-1)^{|\psi|+1}\int \phi_a \wedge(D\psi)_a
\etag\bbb$$
for $\psi,\phi\in\Omega^*(M,\adp)$
implies that
$L$ is antisymmetric under the involution of
$\adp_1\otimes\adp_2$ that exchanges $\adp_1$ and $\adp_2$.
Equivalently, \bbb\ reads
$$ \int <\psi,D\phi>_{\Lie(G)}
= (-1)^{|\psi|+1} \int <\phi,D\psi>_{\Lie(G)}
.\etag\last$$

General elliptic operator theory guarantees that,
as a vector bundle valued form on $M^2$,
$L$ is smooth away from the diagonal $\Delta\subset M\times M$ and has
singularities as one approaches $\Delta$ which are computable by an
explicit local construction.  Further, since all flat bundles are locally
trivial, the singularity must factor as a product of the singularity
for the ordinary exterior derivative times the identity operator on
the Lie algebra.

In fact it turns out that $L$ extends smoothly to a form, $L_B$,
on the differential geometric blowup,
$B_2=BL(M^2,\Delta)$ of $M^2$ along $\Delta$.
$B_2$ is defined by
replacing $\Delta$ by $S(N(\Delta))$, the sphere bundle to the normal
bundle of $\Delta$ in $M^2$.
It comes equipped with a ``blowdown map'' $b:B_2\rightarrow M^2$.
The restriction of $b$ to the interior of $B_2$ is just the identity
map from $M^2\setminus\Delta$ to itself.  The restriction
$\del b$ of $b$ to the boundary of $B_2$ is the bundle projection map
$$ \del b: \del B_2 = S(N(\Delta)) \rightarrow \Delta
.\etag\last$$
This bundle is naturally isomorphic to the bundle
$S(TM)\rightarrow M$.

Abusing notation, we shall denote the bundle $b^*(\adp_i)$ for $i=1,2$
simply by $\adp_i$.
Then $L_B$ belongs to the space $\Omega^2(B_2,\adp_1\otimes\adp_2)$.
Note that on $\del B_2$, $\adp_1=\adp_2$ and
$\adp_1\otimes\adp_2\cong\Hom(\adp_1,\adp_1)$.

We will show in \S\SLT\ that
the restriction of $L_B$ to $\del B_2$ takes the form
$$ L_B|_{\del B_2} = l + (\del b)^*(\rho)
,\etag\last$$
where: (i) $\rho\in\Omega^2(\Delta,\adp_1\otimes\adp_2)$ is smooth,
and (ii)
$l$ factors as a product of
a smooth ordinary form $\lambda\in\Omega^*(S(TM))$ times
the identity in $\Hom(\adp_1,\adp_1)\cong\adp_1\otimes\adp_2$.

The forms $L_B$, $\rho$, and $\lambda$ above are not only smooth, but they
are also closed, as we now show.  First observe that
$$
 \{\Dz,\Dz^{-1}\} = \{\Dz,\dlz\circ\lap_M^{-1}\}
		  = \{\Dz,\dlz\}\circ\lap_M^{-1} -\dlz\circ[\Dz,\lap_M^{-1}]
		  = \id
.\etag\opind$$
Let $D_{M^2}$ denote the exterior covariant derivative operator
on $\Omega^*(M^2,\adp_1\otimes\adp_2)$ (which depends on the choice of $A_0$).
Then the integral kernel version of
\opind\ says that $D_{M^2} L$ is the kernel
for the identity operator, and so is supported on the diagonal.
So, the restriction of $L$ to $M^2\setminus\Delta$ is closed as well as smooth.
Since it's extension $L_B$ to $B_2$ is smooth, it must be closed.
Hence $L_B|_{\del B_2}$ is closed.  However, $\lambda$ is also closed, which
follows from it's explicit description below \singlt.
Therefore, $\rho$ is closed as well.

The natural object that arises from the formulation of superspace perturbation
theory is not the basic propagator $L$, but the
``superpropagator'' $L_s$:
$L_s=s(L)$ is the image of $L$ under the linear map from
$\Omega^2(M^2,\adp_1\otimes\adp_2)$
to $\Omega^2(M^2,\Lambda^2(\adp_1\oplus\adp_2))$
induced by the embedding
$$ s: \adp_1\otimes\adp_2 \rightarrow \Lambda^2(\adp_1\oplus\adp_2)
\etag\super$$
which takes
$\theta_1\otimes\theta_2$ to $\theta_1\wedge\theta_2$.
Similarly, let
$$\rho_s=s(\rho)\in\Omega^2(\Delta,\Lambda^2(\adp_1\oplus\adp_2))
.\etag\last$$

The antisymmetry of $L$ under the involution exchanging
$\adp_1\rightarrow M_1$ and
$\adp_2\rightarrow M_2$
implies that $L_s$ is symmetric under such an involution.
That is,
for $(x_1,x_2)\in M^2$,
$\{j^a_{(1)}\}$ a basis of $\adp_1$, and $\{j^a_{(2)}\}$ a basis of $\adp_2$,
we have
$$ L_s(x_1,x_2) = L_{ab}(x_1,x_2) j^a_{(1)}\wedge j^b_{(2)}
  = - L_{ba}(x_2,x_1) j^a_{(1)}\wedge j^b_{(2)}
  = L_s(x_2,x_1)
.\etag\last$$
This equation implicitly defines an identification of
$\Lambda^*(\adp_1\oplus\adp_2)$ with $\Lambda^*(\adp_2\oplus\adp_1)$.

The Feynman integrands are built up out of the superpropagator $L_s$
as we shall now see.

\bigskip
\newsec\Spert{Formulation of Perturbation Theory}
\medskip

Fix an integer $l\ge 2$, and let $I=3(l-1)$ and $V=2(l-1)$.
Let $M^{\{i\}}$ be the $i^{th}$ copy of $M$ in the Cartesian product
$M^V$ and $\adp_i$ be a copy of $\adp$ over $M^{\{i\}}$.
By abuse of notation,
the pullback of $\adp_i$ by the projection map from $M^V$ to $M^{\{i\}}$
will also be denoted $\adp_i$.
A choice of local trivialization of $\adp_i$ determines sections
$j^a_{(i)}$ of $\adp_i$ corresponding under the trivialization to the
orthonormal basis $\{T_a\}$ chosen for $\Lie(G)$.
Elements of $M^V$ will be written as $\vec x=(x_1,..., x_V)$.

To describe the Feynman amplitude $I_l$ for $l$ loop perturbation
theory, we introduce the bundle
$$A_V^*\equiv\Lambda^*(\adp_1\oplus\adp_2 ...\oplus\adp_V)
\etag\defAV$$
of Grassman algebras over $M^V$.
The fiber of $A_V^*$ at a point is the graded commutative algebra
generated freely by the degree one generators
$\{j^a_{(i)}; i=1,...,V,\; a=1,...,\dim(G)\}$.
The operation of interior product with
the dual basis vector to $j^a_{(i)}$ will be denoted
$\partt{j^a_{(i)}}$; this is a graded derivation of $A_V^*$.

For $i=1,...,V$,
let $Tr_i:A_V^*\rightarrow A_V^*$ be the map
$$\Tr_i\equiv \pi_i\circ
f_{abc}\partt{j^a_{(i)}}\partt{j^b_{i)}}\partt{j^c_{(i)}}
,\etag\defetri$$
where $\pi_i$ is the projection operator onto the subspace of $A_V^*$
of homogeneous degree $0$ element in the $\adp_i$ direction.
The definition of $\Tr_i$ is independent of the choice of trivializations
since $f_{abc}$ is an invariant tensor.  In fact it may be describe more
invariantly as the linear map so that
$$\Tr_i(\theta_1\wedge ... \wedge\theta_n \wedge\omega)=
   \cases{ 0\qquad   & $n \ne 3$ \cr
          -6 <\theta_1, [\theta_2,\theta_3 ]>_{\Lie(G)}\omega \qquad  &$n = 3,$
	  }
\etag\last$$
for $\theta_1,..., \theta_n$ sections of $\adp_i$ and
$\omega$ a section of $A_V^*$ of degree $0$ in the $\adp_i$ directions.

The composition of the $\Tr_i$ acting on an element of
$A_V^*$ produces an element of overall degree $0$ , i.e. a real number.
So acting on forms with values in $A_V^*$, we have a map
$$\Tr^{(V)} \equiv \Tr_1\circ ... \circ\Tr_V:
 \Omega^*(M^V,A_V^*)\rightarrow\Omega^*(M^V)
.\etag\last$$

The Feynman amplitude for
$l$-loop perturbation theory may now be compactly written as
$$I_l(M, A_0, g) \equiv c_{V,I} \int_{M^V} \Tr^{(V)}(L_{tot}{}^I)
,\qquad c_{V,I} = {1\over 2!^I (3!)^V V! I!}
.\etag\defIl$$
The ``total propagator''
$$L_{tot}\in\Omega^2(M^V, A_V^2)\subset\Omega^*(M^V, A_V^*)
$$
will be defined in a moment.
It makes sense to raise $L_{tot}$ to a power since it is
valued in a bundle of algebras.
$L_{tot}{}^I$ has degree $2I=3V$ as a differential form, so that
the integrand in \defIl\ is in fact a top form on $M^V$.
$I_l$ and $L_{tot}$
depend on the flat connection $A_0$ and the metric $g$, since $L$ does.

To define $L_{tot}$, let
$$L_{s,\{i,j\}}\in \Omega^2(M^{\{i,j\}},\Lambda^2(\adp_i\oplus\adp_j))
\qquad\hbox{for $i\ne j$}
$$
be a copy of the superpropagator $L_s$ defined on $M^{\{i,j\}}$ rather
than $M^2$.
The symmetry of $L_s$ under involution means that the definition of
$L_{s,{\{i,j\}}}$ is independent of whether we identify
$M^{\{i,j\}}$ with $M^{\{i\}}\times M^{\{j\}}$ or $M^{\{j\}}\times M^{\{i\}}$.
$L_{s,{\{i,j\}}}$ is smooth away from the diagonal
$\Delta_{\{i,j\}}\subset M^{\{i,j\}}$.
It pulls back via the projection map
$\pi_{\{i,j\}}: M^V\rightarrow M^{\{i,j\}}$ to a form
$$
L_{ab}( x_i, x_j) j^a_{(i)}\wedge j^b_{(j)}= (\pi_{\{i,j\}})^*(L_{s,\{i,j\}})
\in\Omega^2(M^V, A_V^2)
.\etag\notl$$
The pullback operation here is the usual pull back of differential
forms combined with the identification of the pull back of
$\Lambda^*(\adp_i\oplus\adp_j)\rightarrow M^{\{i,j\}}$ with a subbundle
of $\Lambda^*(\adp_1\oplus ...\oplus\adp_V)= A_V^*$.
Since $L_{s,\{i,j\}}$ is smooth away from the diagonal
$\Delta_{\{i,j\}}\subset M^{\{i,j\}}$, the pullback is smooth away from
the diagonal
$$\bar\Delta_{\{i,j\}} = \pi_{\{i,j\}}^{-1}(\Delta_{\{i,j\}}) \subset M^V
.\etag\last$$

For $i=j$, \notl\ seems not to be well defined at
any point in $M^V$ due to the singularity of $L$ near the diagonal.
It can nevertheless be given a sensible interpretation because
 $j^a_{(i)}\wedge j^b_{(i)}$ is anti-symmetric under the exchange
of $a$ and $b$, whereas the singular part of $L$ is symmetric in the Lie
algebra indices.  So we can interpret the singular piece as not making a
contribution and define
$$L_{ab}( x_i, x_i) j^a_{(i)}\wedge j^b_{(i)}
 \equiv \rho_{ab}(x_i,x_i) j^a_{(i)}\wedge j^b_{(i)}
\in\Omega^2(M^V, A_V^2)
.\etag\notm$$
The notation here, as in \notl, is a useful way of
summarizing a complicated pullback.  That is,
\notm\ can also be written as $(f_{\{i\}})^*(\rho_{s,\{i\}})$.
Here $\rho_{s,\{i\}}$ is a copy of $\rho_s$ belonging to
$\Omega^2(M^{\{i\}},\Lambda^2(\adp_i\otimes\adp_i))$ rather than
$\Omega^2(\Delta,\Lambda^2(\adp_1\otimes\adp_2))$ and
$f_{\{i\}}$ is the projection map from $M^V$ to $M^{\{i\}}$.

Finally, $L_{tot}$ is given by
$$ L_{tot} \equiv \Sum_{i,j=1}^V L_{ab}(x_i,x_j) j^a_{(i)}\wedge j^b_{(j)}
.\etag\last$$

\medskip
\noindent{\it Graphical Interpretation:}

To obtain a graphical interpretation of \defIl, we expand
$$ L_{tot}{}^I =\sum_{i_1,j_1 =1}^V ... \sum_{i_I,j_I =1}^V
		\prod_{e=1}^I L_{a_e b_e}(x_{i_e}, x_{j_e})
		  j^{a_e}_{(i_e)} j^{b_e}_{(j_e)}
.\etag\last$$
A choice of $i$'s and $j$'s in the above sum determines a labeled\foot{
  Labelings in [\AS] included an ordering of the edge ends incident on
      any vertex.  It is not necessary to include this in our labelings
      here, since we have not introduced explicit Lie algebra indices in
      our Feynman rules.  Instead, our basic vertex includes a sum over
      orderings of incident edge ends.
  },
oriented graph $\LG$ which has vertices
labeled $1,...,V$, edges labeled $1,...,I$, and has the $e^{th}$ edge
oriented to point from the vertex $i_e$ to the vertex $j_e$
($1\le e\le j$).  In fact, this establishes a one to one correspondence
between the set of individual terms in the above sum and the set of
labeled oriented graphs with Euler characteristic $V-I = 1-l$.
Since $\Tr_i$ vanishes on forms with
degree other than $3$ in the $\adp_i$, only terms corresponding to trivalent
graphs contribute to $I_l$.  Therefore we may write
$$\eqalign{
 I_l &= c_{V,I} \sum_{{\scriptstyle\LG\hbox{ trivalent}\atop
		       \scriptstyle\chi(\LG)=1-l}} I(\LG)
\cr
I(\G)\equiv I(\LG) &\equiv \int_{M^V}\Tr^{(V)}(\cI(\LG))
\cr
\cI(\LG) &\equiv \prod_{e=1}^I L_{a_e b_e}(x_{i_e}, x_{j_e})
		  j^{a_e}_{(i_e)} j^{b_e}_{(j_e)}
.}\etag\graphint$$
We'll refer to  $\cI(\LG)$ as the Feynman integrand and
$I(\G)$ as the Feynman amplitude for $\LG$.
In our notation for $I(\G)$ in \graphint,
we have dropped the underline on $\LG$ since
$I(\G)$ only depends on the topological type
$\G$ of $\LG$, and not on the the labeling.
Although this allows us to equate $I_l$ with a sum over topological types
as is usually done, it will usually be more convenient for us to stick with the
formulation above.

To state our main theorem, we need the amplitude
for connected graphs only:
$$ I_l^{conn} \equiv  c_{V,I}
\sum_{\scriptstyle\LG \hbox{ trivalent}\atop
      \scriptstyle\hbox{connected, $l$ loops}} I(\G)
.\etag\last$$
Since the graphs in the sum above are connected, the Euler characteristic
condition just means that the graphs have $l$ loops.

\bigskip
\newsec\SLT{The Extended Propagator $\tilde L$}
\medskip

In this section we define the extended propagator $\tilde L$
and describe its properties.
The properties of $\tilde L$ will be described first since
that is what is used in the proof of the main Theorem in \S\Stheorem.

\newsubsec\SSPLT{Properties of $\tilde L$}

Let
$\TM$, $\tilde\adp_i$, $\tilde b$, and $\del\tilde b$ be
the bundles $TM\rightarrow M$ and
$\adp_i$ (over whichever base space appropriate), and
the maps $b: B_2\rightarrow M^2$ and
$\del b: \del B_2\rightarrow\Delta$, all trivially crossed with $\Met$.
$\nabla^\TM$ will denote the natural covariant differential on
$\TM\rightarrow M\times\Met$ which is compatible with the inner product on
the fibers.  (At $(z,g)\in M\times\Met$, the inner product is simply $g(z)$.)
See \ntm\ for a more concrete description of $\nabla^\TM$.

The salient features of $\tilde L$ are L1 through L7 below.
L1 -- L3 simply explain what kind of object $\tilde L$ is and how it
is an extension of $L$.  These properties follow immediately from
the definition in \S\SSDLT.  Properties L4 -- L7 concern the nature of
the singularities of $\tilde L$.  They are proved in \S\SSSLT.

\item{L1.}
$\tilde L$ belongs to
$\Omega^2(M^2\times\Met,\tilde\adp_1\otimes\tilde\adp_2)$.

\item{L2.}
Let $\tilde L^{(i)}$ be the piece of $\tilde L$ of homogeneous form degree
$i$ in the $\Met$ directions.  Then
$\tilde L^{(0)}$ equals the basic propagator $L$ (considered as a function
on $\Met$).

\item{L3.}
$\tilde L$ is smooth and covariantly closed away from $\Delta\times\Met$.

\item{L4.}
The restriction of $\tilde L$ to $[M^2\setminus\Delta]\times\Met$
extends smoothly to a covariantly closed form
$$\tilde L_B \in\Omega^2(B_2\times\Met,\tilde\adp_{1}\otimes\tilde\adp_{2})
.\etag\last$$

\item{L5.}
There are smooth closed forms
$$\eqalign{
\tilde\rho\in &\Omega^2(\Delta\times\Met,\tilde\adp_1\otimes\tilde\adp_2),
\hbox{ and} \cr
\tilde l \in &\Omega^2(\del B_2\times\Met,\tilde\adp_{1}\otimes\tilde\adp_{2})
}$$
so that
$$
\tilde L_B|_{\del B_2\times\Met} = \tilde l + (\del\tilde b_2)^*(\tilde\rho)
.$$

\item{L6.}
$\tilde l$ factors into a ``manifold piece'' times a ``Lie algebra piece'',
$$\eqalign{
 \tilde l = &\tilde\lambda\otimes \id_\adp
 \cr
 \tilde \lambda \in &\Omega^2(\del B_2\times\Met)
 \cr
 \id_\adp\in &\Gamma(\del B_2\times\Met,\tilde\adp_{1}\otimes\tilde\adp_{2})
.}\etag\last$$
$\id_\adp$ is the inverse to the invariant metric on $\Lie(G)$ made
into a bundle section.  Under the identification
$$\tilde\adp_1\otimes\tilde\adp_2|_{\del B_2}
  = \Hom(\tilde\adp_1,\tilde\adp_1)
,\etag\last$$
$\id_\adp$ is the identity element on each fiber of
$\tilde\adp_1|_{\del B_2}$.

\item{L7.}
Identifying $\del B_2\rightarrow\Delta$ with
$S(TM)\rightarrow M$, $\tilde\lambda$ may be viewed as an element
of $\Omega^2(S(TM)\times\Met)$.  As such, it is given by the following
local, universal formula involving the covariant differential
$\nabla^\TM$ and its curvature $2$-form
$\tilde\BOmega\in\Omega^2(M\times\Met,\Hom(\TM,\TM))$.
$$\eqalign{
\tilde\lambda((z,\hat u),g) =
-{1\over 8\pi}  & {\det(g_{pq}(z))}^\half\epsilon_{ijk} (\hat u^i)
\cr
&\times \left[(d_{vert}\hat u^j)(d_{vert}\hat u^k) +
        \tilde\Omega^j{}_l(z,g) g^{lk}(z)
\right]
.}\ftag\singlt$$

\medskip

In \singlt, $\hat u\in S(TM)|_z$ is a vector in $T_zM$ of unit length
with respect to the inner product $g(z)$.  \singlt\ is written using
coordinates
$\{z^i\}$ about $z\in M$ and the components
$\{\hat u^i\}$ for $\hat u=\hat u^i\partt{z^i}$.
$d_{vert}\hat u^i$ is the projection of $d\hat u^i$ onto the space of
verticle $1$-forms determined by $\nabla^\TM$.

Let $\{\partt{z^i}\}$ be the local trivialization of $\TM$ associated
to the coordinates $\{z^i\}$. $\nabla^\TM$ is given by
$$\eqalign{
[\nabla^\TM_{\part{z^i}}\,\partt{z^k}]{(z,g)} =& \Gamma_{ik}^j(z)\partt{z^j}
\cr
[\nabla^\TM_m\partt{z^k}]{(z,g)} = & \half g^{jl}(z) m_{lk}(z) \partt{z^j}
\qquad\hbox{for $m\in T_g\Met=\Gamma(\Sym^2(TM)\rightarrow M)$}
.}\ftag\ntm$$
Here $\{\Gamma_{ik}^j\}$ are the Christoffel symbols for the metric
connection determined by $g$.
The vertical projection of the function $u^i$ of a vector $(z,u)$ in $TM$
is
$$ d_{vert} u^j = d u^j +
   [\Gamma_{ik}^j dz^i+\half(g^{-1}\delta g)^j{}_k]_z u^k
.\etag\dvu$$
$d_{vert}\hat u^i$ in \singlt\ is the value at $(z,\hat u)$ of the pullback
of $d_{vert} u^j$ by the inclusion map $S(TM)\hookrightarrow TM$.

$\nabla^\TM$ can be described invariantly.
Equip $M\times\Met$ with a Riemannian metric of the following form
$$ <(v_1,m_1), (v_2,m_2)>_{(x,g)} = g_x(v_1,v_2) + G_g(m_1,m_2)
,\etag\rmomm$$
for $(x,g)\in M\times\Met$, $v_1,v_2\in TM_x$, $m_1,m_2\in T\Met_g$.
$G$ is {\it any} Riemannian metric on $\Met$
(not necessarily a natural one).
So $\TM$ is the subbundle of $T(M\times\Met)$ whose orthogonal
complement is $M\times T\Met$.
Then $\nabla^\TM$ is the covariant differential on $M\times\Met$
followed by the projection operator $\pi_{\TM}$ onto  $\TM$,
i.e.
$$ \left[\nabla^\TM_{(v,m)} w\right](x,g)
 = \left[\pi_{\TM}\circ \nabla^{T(M\times\Met)}w\right](x,g)
,\etag\last$$
for $w$ a section of $\TM$.
We leave to the reader to check that this does give the connection above
and to compute the curvature formulas in the next paragraph.

The curvature two form of $\nabla^\TM$ decomposes as a sum
$$\tilde\BOmega
=\tilde\BOmega^{(2,0)}+\tilde\BOmega^{(1,1)}+\tilde\BOmega^{(0,2)}
,\etag\expom$$
where
$\tilde\BOmega^{(i,2-i)}$ has form degree $i$ in the $M$ directions and
$2-i$ in the $\Met$ directions.  From \ntm, it follows that
\eqna\eto
$$\eqalignno{
[\tilde\BOmega^{(2,0)}(z,g)]^k{}_l =&
     \left[
     \partt{z_i}\Gamma_{jl}^k +\Gamma_{im}^k\Gamma_{jl}^m
     \right]_z dz^i \wedge dz^j
&\eto{.1}\cr
[\tilde\BOmega^{(1,1)}(z,g)]^k{}_l =&
     \left[
     \delta\Gamma_{il}^k
     - \half \nabla_{\part{z^i}} (g^{-1}\delta g)^k{}_l
     \right]_z \wedge dz^i
&\eto{.2}\cr
[\tilde\BOmega^{(0,2)}(z,g)]^k{}_l =&
     -\quarter[(g^{-1}\delta g)^k{}_n \wedge (g^{-1}\delta g)^n{}_l]_z
.&\eto{.3}
}$$
Here $\delta\Gamma_{il}^k(z)$ and $\delta g_{ml}(z)$ are the
exterior derivatives in the metric directions of the functions
$\Gamma_{il}^k(z)$ and $g_{ml}(z)$.
The covariant derivative operator in \eto{.2}\ acts on the indices
$k$ and $l$.  This comes from the commutator of the right hand sides of
the two equations in \ntm.  Note that
\eto{.1}\ equals the usual Riemannian curvature $\BOmega^k{}_l(z)$,
considered as a function on $\Met$.  One check that the relative coefficients
in \eto{.2}\ are correct is that the sum of the two terms
is antisymmetric in $k$ and $l$.

\newsubsec\SSDLT{Definition of $\tilde L$ and Proof of L1-L3}

Let $W$ be the vector bundle $\Lambda^*(T^*(M\times\Met))\otimes\adp$
over $M\times\Met$.
For $g\in\Met$, let $\tilde W_g  = \Gamma(M, W|_{M\times\{g\}})$.
This may be identified with a graded tensor product
$\tilde W_g = \Omega^*(M,\adp) \hat\otimes \Lambda^*(T^*\Met_g)$.
$\tilde W_g$ is the fiber at $g$ of a vector bundle $\tilde W\rightarrow\Met$.
So
$\Gamma(\Met, \tilde W) = \Omega^*(M\times\Met,\adp)$.
(This may be viewed as a definition of what is meant by sections of the
bundle $\tilde W$ whose fibers are infinite dimensional.)

Let $D_{M\times\Met}$ be the covariant exterior derivative operator on
$\Omega^*(M\times\Met,\adp)$, and $\tl D_{M}$, $\tl D_{\Met}$
it's pieces in the indicated directions.
$\tl D_M$ can be viewed as the operator $D=D_M$ on $\Omega^*(M,\tilde\adp)$,
made to act on the sections of $\tilde W$ through its action on each fiber
separately.
The action
$(\tl D_M)_g=D_M\hat\otimes\id_{\Lambda^*(T^*\Met_g)}$ on $\tilde W_g$
will be abbreviated simply by $D_M$.
Let $\kappa:\Omega^*(M,\adp)\rightarrow\Omega^*(M,\adp)$ be the operator
$\kappa\omega= (-1)^p\omega$ for $\omega\in\Omega^p(M,\adp)$.
The operators $\dlz$, Hodge star $*$, and $\kappa$ determine operators
$\tilde\dlz$, $\tilde *$, and $\tilde\kappa$ on $\Omega^*(M\times\Met,\adp)$
which are related by $\tilde\dlz=\tilde *\tilde D_M\tilde *\tilde\kappa$.

Define
$$\tflap\equiv\{D_{M\times\Met}, \tl \dlz\}:
\Omega^*(M\times\Met,\tilde\adp)\rightarrow\Omega^*(M\times\Met,\tilde\adp)
.\etag\last$$
Then
$$\tflap =\tl\lap_M+ \tilde A,\hbox{ where}\quad
\tilde A = \{\tl *\{\tl D_{\Met},\tl *\},\tl\dlz\}
.\etag\meo$$
Notice that $\tl\lap_M$
($\lap_M$ acting on $\Omega^*(M\times\Met,\tilde\adp)$)
is a second order elliptic operator in the $M$ directions,
$\tl A$ is a first order operator in the $M$ directions,
and $\tl\lap_M$ and $\tl A$ both involve no derivatives in
the $\Met$ directions.
So $\tflap$ is an operator on $\Gamma(\Met,\tilde W)$ which acts on each fiber
of $\tilde W$ separately.  On $\tilde W_g$, it acts by the elliptic operator
$$\tflap_g = \lap_M + \tl A_g
.\etag\llg$$
Since
$\lap_M$ is invertible and $\tl A_g$ increases form degree by $1$ in the $\Met$
directions, $\tflap_g$ is also invertible.
$\tflap^{-1}$ is the operator on sections of $\tilde W$ coming from
the action $(\tflap_g)^{-1}$ on the fiber $\tilde W_g$ for $g\in\Met$.

The extended propagator
$\tilde L\in\Omega^*(M_1\times M_2\times\Met,\tilde\adp_1\otimes\tilde\adp_2)$
 is the integral kernel for the operator
$$\tl \dlz \circ \tflap^{-1}:
 \Omega^*(M\times\Met,\adp)\rightarrow\Omega^{*-1}(M\times\Met,\adp)
.$$
This means that
$$(\tl\dlz\circ O^{-1}\tilde\psi)_a(x,g)
  = \int_{y\in M_2} \tilde L_{ab}(x,y,g)\wedge\tilde\psi_b(y,g)
\quad\hbox{for $\tilde\psi\in\Omega^*(M\times\Met,\tilde\adp)$}
\etag\last$$
or, equivalently, that
$$ \left(\dlz\circ(\tflap_g)^{-1}\psi\right)_a(x)=
 \int_{y\in M_2} \tilde L_{ab}(x,y,g)\wedge\psi_b(y)
\quad\hbox{for $g\in\Met$, $\psi\in\tilde W_g$}
.\etag\iktl$$

To describe $\tilde L$ more explicitly, let
$\tilde G\in\Omega^3(M_1\times M_2\times\Met,\tilde\adp_1\otimes\tilde\adp_2)$
be the integral kernel for $\tflap^{-1}$, defined by
$$ \left((\tflap_g)^{-1}\psi\right)(x)
    =\int_{y\in M_2} \tilde G(x,y,g)\wedge\psi(y)
\quad\hbox{for $g\in\Met$, $\psi\in\tilde W_g$}
.\etag\iktk$$
For fixed $g\in\Met$, $\tilde G(\cdot,\cdot, g)$
is the integral kernel for $(\tflap_g)^{-1}$.  The
Hadamard paramatrix construction for $\tflap_g$ shows that
$\tilde G$ is smooth
away from the diagonal and gives an explicit prescription for calculating
its singularities near the diagonal.  The fact that $\tilde G$ is
smooth in $g$ also follows from the general construction.
Thus,
$$ \tilde L(x,y,g) = - \dlz_x \tilde G(x,y,g)
\etag\rellk$$
is smooth in $x$, $y$, and $g$ away from points with $x=y$.
In \rellk, $\dlz_x$ is the operator $\dlz$ acting in the directions
along $M_1$ to which the point $x$ belongs.

\medskip

We now prove property L2 of $\tilde L$.
Choose $g\in\Met$ and $\psi\in\Omega^*(M,\adp)$ and
identify $\Omega^*(M,\adp)$ with the subspace
$\Omega^*(M,\adp)\otimes\Lambda^0(T^*\Met_g)$ of $\tilde W_g$.
Let $\eta=(\tflap_g)^{-1}\psi$ and
$\eta_k$ be the piece of $\eta$ of degree $k$ in the $\Met$
directions.
So
$$\eqalign{
\lap_M\eta_0 &=\psi
\cr
\lap_M\eta_k &= - \tilde A_g\eta_{k-1},\qquad\hbox{for $k>1$.}
}\etag\last$$
Hence $\eta_0 =\lap_M^{-1}\psi$
and $\dlz\eta_0$ equals both $\dlz\circ\lap_M^{-1}\psi$ and
the piece of $\dlz\circ(\tflap_g)^{-1}\psi$
of degree $0$ in the $\Met$ directions.
This means that
$$\int_{y\in M_2}\tilde L^{(0)}(x,y,g)\wedge \psi(y)
   = \int_{y\in M_2} L(x,y;g) \wedge\psi(y)
$$
for each $x\in M$, $g\in\Met$ and $\psi\in\Omega^*(M)$.
The preceeding statement says exactly that
$\tilde L^{(0)}$ equals $L$.

Property L3  follows by generalizing \opind.
First observe that
$\{D_{M\times\Met},\tflap\}=0$ and so $\{D_{M\times\Met},\tflap^{-1}\}=0$.
Therefore
$$\{D_{M\times\Met},\tilde\dlz\circ\tflap^{-1}\}
  = \{D_{M\times\Met},\tilde\dlz\}\tflap^{-1}
  =\id_{\Omega^*(M\times\Met,\tilde\adp)}
.\etag\last$$
Hence $D_{M^2\times\Met}\tilde L$ is the integral kernel for the identity
operator, and so vanishes away from $\Delta\times\Met$.

\newsubsec\SSSLT{The Extension $\tilde L_B$ of $\tilde L$}

To prove the extension
$\tilde L_B\in\Omega^2(B_2\times\Met,\tilde\adp_1\otimes\tilde\adp_2)$
exists and satisfies properties L4-L7, we need to calculate the
singularity near $\Delta\times\Met$ of $\tilde L$.
We shall use a version of the rescaling used by Getzler [\GET] in
studying the heat kernel for generalized Laplacians to prove the
local index theorem.  See also [\BGV].

Our proof will be rather condensed.  Further elaboration, generalization,
and discussion of the relation with heat kernels can be found in a
forthcoming paper by the first named author [\AXE].
In particular, it will be shown that the restriction as a form
of $L_B$ to $\del B_2$ may be derived from the equivariant Thom
class obtained as a scaling of the heat kernel singularity in
[\MATHAI].

Throughout the discussion the metric $g\in\Met$ will be fixed.
The space $\Lambda^*(T^*\Met_g)$  will be abbreviated as $\lmg$,
and we write $\flap$ for $\tflap_g$, $\bar G$ for the integral
kernel for $\flap$, and $\bar L$ for the integral kernel for
$\dlz\circ\flap^{-1}$.  So $\bar G(x,y) =\tilde G(x,y,g)$,
$\bar L(x,y)=\tilde L(x,y,g)$.

Since propagator singularity calculations are local and the flat
connection $A_0$ is locally trivial, it is automatic that the singularity
factorizes into a manifold piece (independent of $A_0$) times the
identity operator on $\adp$.  Therefore we may specialize to the case
when the group $G$ has one element.

\bigskip

\noindent{\it Coordinates, Taylor series, and Singular Series}

To describe the singularity calculation we need to describe
coordinates on $M_1\times M_2$ near $\Delta$, several gradings of
the space of $\lmg$ valued forms defined near $\Delta$,
and several ways to package generalized ``Taylor'' series near $\Delta$ for
such forms and operators acting on them.

Choose $\epsilon>0$ much smaller than the injectivity radius of $M$ and
let $N=\{(z,u)\in TM; ||u||<\epsilon\}$ be the open ball of
radius $\epsilon$ in $TM$.
Let $E:N\rightarrow M_1\times M_2$ be the map sending
$(z,u)$ to $(x,y)=E(z,u)\equiv (\exp_z u, \exp_z -u)$.
$E$ is a diffeomorphism of $N$ onto a neighborhood of $\Delta$ in $M^2$.
The restriction $E'$ of $E$ to
$N'\equiv\{(z,u)\in N; u\ne 0\}$ is a diffeomorphism onto
$E(N)\setminus\Delta$.

Given local coordinates $\{z^i\}$ on a open set $U$ in $M$,
define local coordinates on $N\intersect TU$ by taking
the coordinate of the point $(z,u)$ to be $(z^i, u^i)$, where $(z^i)$ are
the coordinates of $z$ and $u= u^i\part{z^i}|_z$.

Let $S=u^i\part{u^i}$ be the vector field on $TM$ generating
dilation.  In local coordinates $\cL_S$ acts on $\Omega^*(N,\lmg)$ by
$$\cL_S = u^i\part{u^i} + e(du^i) i(\part{u^i})
.\etag\scales$$
Given $\omega\in\Omega^*(N',\lmg)$,
we say that $\omega$ has total degree $|\omega|_{tot}$
if $\cL_S\omega = |\omega|_{tot}\omega$.
Similarly, we say that $\omega$ has degree $|\omega|_u$ in $u$
if $u^i\part{u^i}\omega=|\omega|_u \omega$.
Finally, we say that
$\omega$ has degree $|\omega|_{du}$ in $du$
if $e(du^i) i(\part{u^i})\omega =|\omega|_{du}\omega$, i.e. if
$\omega$ has form degree $|\omega|_{du}$ in the $u^i$ directions.
Equation \scales\ says that the total degree of $\omega$ equals
the degree in $u$ plus the degree in $du$.

Note that the notion of $u$ degree and $du$ degree depend on the choice
of coordinates $z^i$.  Properly speaking we should only talk about
degree in $u$ and $du$ of a form on
the subset of $N$ where the coordinates $\{z^i,u^i\}$ are defined.
We will not introduce any special notation for this, however, since
the final results below for the propagator singularities graded by total
degree are coordinate system independent.
Alternatively, we could introduce covariant notions of $u$ degree and $du$
degree.

Suppose given smooth $\phi\in\Omega^*(M_1\times M_2\setminus\Delta,\lmg)$
and smooth $\phi_s\in\Omega^*(N',\lmg)$ for $s=s_0, s_0+1,...$\ .
We say that
$\sum_{s=s_0}^\infty \phi_s$ is a {\it singular series} for $\phi$ if
for any $k$, there is a $K_0$ so that, whenever $K\ge K_0$,
$E'^*(\phi)-\sum_{s=s_0}^K \phi_s$ extends
$k$ times continuously differentiably across the zero section
(i.e. to all of $N$).
If $|\phi_s|_{tot}=s$ (resp. $|\phi_s|_u=s$) for all $s$,
we say that $\phi_s$ is the singularity of $\phi$
of total degree (resp. degree in $u$) $s$.  Note that the singularity of
$\phi$ of a given
degree is unique up to addition of a form smooth on all of $N$.

The total degree, degree in $u$, and degree in $du$ of a differential operator
$P$ on $\Omega^*(N,\lmg)$ is the amount by which it shifts the respective
notions of degree, e.g.
$$|P\omega|_{tot}= |P|_{tot} + |\omega|_{tot}
.$$
Suppose $Q$ is an order $\ord(Q)$ differential operator acting on
$\tilde W_g=\Omega^*(M,\lmg)$ with smooth coefficients.
Let $Q_x$ be the differential operator on
$\Omega^*(M_1\times M_2,\lmg)$ so that
$Q_x (\omega_1(x)\wedge\omega_2(y)) = (Q_x\omega_1(x))\wedge\omega_2(y)$
for $\omega_1\in\Omega^*(M_1,\lmg)$, $\omega_2\in\Omega^*(M_2)$.
$Q_x$ has a Taylor series expansion which can be described as
follows.
Let $E^*(Q_x)$ be the pull-back of $Q_x$ to a differential operator
on $\Omega^*(N,\lmg)$.
In local coordinates
$$E^*(Q_x) = \sum_{I,J;\atop |I|+|J|\le\ord(Q)}
Q^{I,J}(z,u)\part{z^I}\part{u^J}
,\etag\last$$
where $I=(i_1,...,i_k)$ and $J=(j_1,...,j_l)$ are multi-indices,
$|I|=k$, $|J|=l$, $\part{z^I}=\part{z^{i_1}}...\part{z^{i_l}}$,
$\part{u^J}=\part{u^{j_1}}...\part{u^{j_l}}$, and
$Q^{I,J}(z,u)$ is a linear transformation of
$\Lambda^*(T^*(TM)_{(z,u)})\hat\otimes\lmg$ depending smoothly on $z$ and $u$.
Let $Q^{I,J}(z,u)_{(k)}$
be the $k^{th}$ order term in the Taylor expansion of
$Q^{I,J}(z,u)$ in the variable $u$.
Set
$$(Q_x)_{(p)} = \sum_{I,J,k;\atop k-|J| =p} Q^{I,J}(z,u)_{(k)}
\part{z^I}\part{u^J}
.\etag\last$$
This vanishes unless $p\ge -\ord(Q)$.
We call
$\sum_p (Q_x)_{(p)}$ the Taylor series expansion of $Q_x$ by degree in $u$
for the following reason.
If $\phi_{(p)}$ is the singularity of $\phi$ of degree $p$ in $u$,
then
$$ (Q_x\phi)_{(n)}\equiv \sum_{p,q; p+q=n} (Q_x)_{(p)}\phi_{(q)}
\etag\dts$$
is the singularity of $Q_x\phi$ of degree $n$ in $u$.

The Taylor series for $Q_x$ may be further refined by writing
$$(Q_x)_{(p)} = \sum_q (Q_x)_{(p,q)}
\etag\last$$
where $(Q_x)_{(p,q)}$ is the piece of $(Q_x)_{(p)}$ which shifts
$du$ degree by $q$.
Also define
$$ (Q_x)_{[s]} \equiv \sum_{p,q;\atop p+q=s} (Q_x)_{(p,q)}
.\etag\last$$
Then $\sum_s (Q_x)_{[s]}$ is the Taylor series expansion of $Q_x$
by total degree; it obeys an equation similar to \dts\ but with
degree in $u$ replaced by total degree.
In summary,
$$\eqalign{
|(Q_x)_{(p)}|_u &= p, \qquad |(Q_x)_{[s]}|_{tot} =s
\cr
(Q_x)_{(p,q)}|_u &= p,  \qquad |(Q_x)_{(p,q)}|_{du} = q
.}\etag\last$$

It is easy to see that the leading terms in the Taylor expansions of
$\flap_x$ and $\dlz_x$ by total degree are
$(\flap_x)_{[-2]}$ and $(\dlz_x)_{[-2]}$, respectively.
In other words
$(\flap_x)_{(p,q)}$ and $(\dlz_z)_{(p,q)}$ vanish for $p+q<-2$.
Straightforward calculation yields
\eqna\sings
$$\eqalignno{
4(\flap_x)_{[-2]} &= - g^{ij}(z) X_i X_j
            + g^{jk}(z) \tilde\Omega(z,g)^i{}_k\> i(\part{u^i})\, i(\part{u^j})
&\sings{.1}\cr
4(\dlz_x)_{[-2]} &= - g^{ij}(z)\, i(\part{u^i}) X_j
,&\sings{.2}
}$$
where
$$\eqalign{
X_k &=\part{u^k}
  -\left[\Gamma_{ik}^j(z) dz^i + \half (g^{-1}\delta g)^j{}_k\right]
   \, i(\part{u^j})
}\etag\xy$$
and
$\tilde\Omega(z,g)$, $g^{-1}\delta g$ are as described in
\eto{}\ and in what follows.
The leading singularity in the expansion of $(\flap_x)$
by degree in $u$ is
$$ (\flap_x)_{(-2)} = (\flap_x)_{(-2,0)} =
   -{1\over 4} g^{ij}(z)\part{u^i}\part{u^j}
.\etag\ldfu$$

\noindent{\it Singularity of $\bar G$ and $\bar L$}

The Hadamard parametrix construction [\Ho] applied to the
elliptic operator $\tflap_g$ determines a singular series
$\sum_{p=-1}^\infty \bar G_{(p)}$ for $\bar G$ where $|\bar G_{(p)}|_u=p$.
The series is constructed so that
$\bar G_{(p)}$ is of the form $||u||^{-1} \bar F_{p+1}$, where
$||u||= g_z(u,u)^{\half}$ and
$\bar F_{p+1}\in\Omega^*(N,\lmg)$ depends smoothly on $z$ and
is a polynomial of degree $p+1$ in its dependence on $u$.
(Hadamard's construction uses the map
$(z,u)\mapsto (z,\exp_z u)$ rather than $E$, but
the results immediately translate into the packaging used here.)

The leading singularity $\bar G_{(-1)}$  is
$$\bar G_{(-1)}\equiv {1\over 24\pi ||u||}
     \sqrt{\det(g_{lm}(z))}\epsilon_{ijk} du^i du^j du^k
.\etag\tgzer$$
$\bar G_{(p)}$ is then determined inductively in $p$ from the fact that
$\flap_x\bar G(x,y)=0$ for $(x,y)$ away from $\Delta$.  The singular
piece of this equation of degree $n$ in $u$ is
$\sum_{k+l=n} (\flap_x)_{(k)}\bar G_{(l)} =0$.
In other words
\eqna\indg
$$\eqalignno{
(\flap_x)_{(-2)}\bar G_{(-1)} & =0            &\indg{.1}
\cr
(\flap_x)_{(-2)}\bar G_{(p)} &
  = \sum_{-1\le l\le p-1} (\flap_x)_{(-2+p-l)}\bar G_{(l)}
 \quad\hbox{for $p\ge 0$.}			 &\indg{.2}
}$$
\indg{.1}\ follows because $G_{(-1)}$ is the flat space propagator.
Equation \indg{.2}\ is an algebraic equation for the polynomial
$\bar F_{p+1}$. Ellipticity of $\flap$ implies that this equation
has a unique solution.\foot{
For a general elliptic operator, the $F$'s might also depend on
powers of $\ln(||u||)$.
No such powers appear here because, using a covariant grading rather
than the coordinate dependent grading, $(\flap_x)_{(-1)}$ vanishes.
   }

\bigskip

Let $\bar G_{(p,r)}$ be the piece of $\bar G_{(p)}$ of degree $r$ in
$du$.
The piece of \indg{.2}\ of degree $r$ in $du$ is
$$(\flap_x)_{(-2,0)} \bar G_{(p,r)} =
 \sum_{-1\le\l\le p-1} \sum_q (\flap_x)_{(-2+p-l,q)}\bar G_{(l,r-q)}
\quad\hbox{for $p\ge 0, 0\le r\le 3$.}
\etag\gpr$$
Now we show that $\bar G_{(p,r)}=0$ for $p+r<2$ by induction on $p$.
For $p=-1$, the result follows from \tgzer.
For $p\ge 0$, $p+r<2$, it suffices to show that the right hand side of \gpr\
vanishes (since $\bar G_{(p,q)}$ is determined uniquely by \gpr).
This follows since $p+r<2$ implies either
$-2+p-l+q<-2$, and so $\flap_{(-2+p-l,q)}=0$, or else
$l+r-q<2$, and so $\bar G_{l,r-q}=0$ by the inductive hypothesis.

Let
$\bar G_{[s]}\equiv\sum_{p,q; p+q=s} \bar G_{(p,q)}$
be the piece of the singularity of $\bar G$ of total degree $s$.
The result of the last paragraph says that $\bar G_{[s]}$ vanishes for $s<2$.
Equations \tgzer\ and \gpr\ imply that $\{\bar G_{[s]},s\ge 2\}$ is
uniquely determined by the conditions that:
\item{U1.}
$\bar G_{(-1,3)}$, the piece of $\bar G_{[2]}$ of degree $3$ in $du$,
is given by the right hand side of \tgzer.
\item{U2.}
$||u||\bar G_{[s]}(z,u)$ is a polynomial in its dependence on $u$.
\item{U3.}
Away from $u=0$,
\eqna\bgtot
$$\eqalignno{
(\flap_x)_{[-2]}\bar G_{[2]} &=0
&\bgtot{.1}\cr
(\flap_x)_{[-2]} \bar G_{[s]}&=
   \sum_{2\le t< s} (\flap_x)_{[-2+s-t]}\bar G_{[t]}
\quad\hbox{for $s>2$.}
&\bgtot{.2}
}$$

\medskip
We shall only have need of the explicit formula for $\bar G_{[2]}$:
$$
\bar G_{[2]}(z,u) =
      {1\over 24\pi ||u||}\sqrt{\det(g_{lm}(z)} \epsilon_{ijk}
 \left[ d_{vert} u^i d_{vert} u^j d_{vert} u^k
 	-3||u||^2\tilde\Omega^i{}_k g^{kj} d_{vert} u^k \right]
.\etag\expgt$$
Since the right hand side obviously satisfies U1 and U2,
one need only check \bgtot{.1}\ to verify \expgt.
This follows by plugging \expgt\ and \sings{.1}\ into \bgtot{.1}\
and calculating.

Since $\bar L(x,y)= -\dlz_x \bar G(x,y)$, $\bar L$ has a
singular series graded by total degree of the form
$\sum_{s=0}^\infty \bar L_{[s]}$, where
$$ \bar L_{[s]} = -\sum_{-2\le t\le s-2} (\dlz_x)_{[t]} (\bar G)_{[s-t]}
.\etag\serl$$
Furthermore
$||u||^3 \bar L_{[s]}$ depends polynomially on $u$.  This follows because
$\ord(\dlz)=1$ and $||u||\bar G_{[s]}$ is a polynomial in $u$.

Using \sings{.2}\ and \expgt\ to evaluate \serl\ for $s=0$, we find
$$\eqalign{
 \bar L_{[0]} = -(\dlz_x)_{[-2]} (\bar G)_{[2]}
 = -{1\over 8\pi}  &{\det(g_{pq}(z))}^\half\epsilon_{ijk} (\hat u^i)
\cr
&\times \left[(d_{vert}\hat u^j)(d_{vert}\hat u^k) +
               \tilde\Omega^j{}_l(z,g) g^{lk}(z) \right]
,}\etag\sngl$$
where $\hat u = u/||u||$.
This has exactly the same form as the right hand side of \singlt.

\medskip
\noindent{\it Extension to $B_2$}

Identify $N'$ with
$S(TM)\times(0,\epsilon)$ via the map
$$ N'\ni (z,u) \mapsto ((z,\hat u),||u||) \in S(TM)\times (0,\epsilon)
.\etag\last$$
Let $E_B:S(TM)\times [0,\epsilon)\rightarrow B_2$ be the map
$$ ((z,\hat u),r) \mapsto\cases{
     (z,\hat u) \in \del B_2
& $r=0$ \cr\casespace
     E(z,r\hat u)\in M^2\setminus\Delta = B^2\setminus\del B_2
& $r>0$.}
\etag\last$$
$E_B$ is a diffeomorphism onto an open neighborhood of $\del B_2$
(by definition of the differentiable structure on $B_2$).
The restriction of $E_B$ to $N'\cong S(TM)\times (0,\epsilon)$
agrees with $E'$; and $E_B|_{S(TM)\times\{0\}}$ is a diffeomeomorphism
of $S(TM)\times\{0\}$ with $\del B_2$.

Observe that
$$\bar L_{[s]}=\cases{
  ||u||^s D_s 					&$s=0$
\cr
  ||u||^s D_s + ||u||^{s-1} d(||u||) E_{s-1}   &$s>0$
}\etag\last$$
where $D_s$ and $E_{s-1}$ are polynomials in
$\hat u^i$ and $d\hat u^i$ (whose coefficients are smooth forms in $z$)
of degree $s$ and $s-1$ respectively.
So $\bar L_{[s]}$ extends smoothly to $S(TM)\times[0,\epsilon)$;
$\bar L_{[s]}|_{S(TM)\times\{0\}}$ vanishes for $s>0$; and
$\bar L_{[0]}|_{S(TM)\times\{0\}}$ is given by the right hand side of
\singlt.

That $\sum_{s=0}^\infty\bar L_{[s]}$ is a singular series for $\bar L$
means that there are
forms $\bar\rho_K\in\Omega^*(E(N),\lmg)$ which become arbitrarily
differentiable for large $K$ so that
$$(E')^*(\bar L) = (E')^*(\bar\rho_K) +\sum_{s=0}^K L_{[s]}
.\etag\last$$
This implies that
$\bar\rho\equiv(\bar\rho_K)|_\Delta$ is independent of $K$ and hence smooth.
$\bar\rho$ is the restriction (as a bundle section) of a smooth form
$\tilde\rho\in\Omega^2(\Delta\times\Met)$ to $\Delta\times\{g\}$.
By the results of the last paragraph,
$(E')^*(\bar L)$ extends smoothly to $S(TM)\times [0,\epsilon)$ and has
restriction to $\del B_2= S(TM)\times \{0\}$ equal to
$$[\tilde\lambda +(\del\tilde b_2)^*(\tilde\rho)]|_{\del B_2\times\{g\}}
\etag\bdrv$$

Using $E_B$ to identify $S(TM)\times[0,\epsilon)$ with a neighborhood
of $\del B_2$ in $B_2$ and using the smoothness of $\bar L$ and $\bar L_{[s]}$
in their dependence on $\Met$, it follows that $\tilde L$ extends to
a smooth form $\tilde L_B\in\Omega^2(B_2\times\Met)$ whose
restriction to $\del B_2\times\Met$ is
$\tilde\lambda+(\del\tilde b_2)^*(\tilde\rho)$.
Since we have already shown that $\tilde L$ is closed and direct
calculation shows $\tilde\lambda$ is closed, it follows that $\tilde\rho$
and $\tilde L_B$ are closed.

We have now shown L4-L7 when the group $G$ is a point.
For general $G$ the only change needed in the above
discussion is that all forms become $\adp_1\otimes\adp_2$
valued and the singularity $\bar L_{[s]}$ gets multiplied by
(the pullback by $E$ of) the $\adp_1\otimes\adp_2=\Hom(\adp_2,\adp_1)$
valued tensor whose value at $(x,y)\in E(N)$ is the
parallel transport homomomorphism along the short geodesic from $x$ to $y$.

\newsec\SBV{The Compactification $M[V]$}
\medskip

In this section we will define a compactification
$M[V]$ of
$$M^V_0\equiv M^V\setminus\union\bar \Delta_{\{i,j\}}
$$
and describe some of its properties.
As mentioned in the introduction, $M[V]$ is a manifold with corners.
That is, it is locally modeled on the space
$C_n\equiv\{(t_1, ..., t_n)\in \IR^n; t_i\ge 0\} $
(where $n=\dim(M[V])= 3V$) with smooth overlap maps.
Smooth maps between open sets in $C_n$ are maps that extend
smoothly to open neighborhoods in $\IR^n$.
We will denote by $\del_k M[V]$ the ``codimension $k$ boundary''
of $M[V]$, that is, the points in $M[V]$ with at least $k$ coordinates
vanishing.  So $\del M[V]=\del_1 M[V]$ is the full boundary.
$\del_k M[V]$ is not a manifold,
but $\del_k M[V]\setminus\del_{k+1} M[V]$ is a disjoint union of smooth
pieces, the codimension $k$ open strata, as we shall see.
The reason $\del_k M[V]$ is not
smooth is that the closed codimension $k$ strata have common boundaries.
(Think of the edges of a cube, which are the intersections of the face;
 or the vertices of a cube, which are the intersections of the edges.)

There are several equivalent definitions of $M[V]$ which can be made by
taking the definitions in the algebro-geometric context of [\FM] and
replacing algebraic geometeric blowups with differential geometric
blowups, i.e. replacing  projective spaces by spheres.
We will not give a complete treatment extending [\FM] to the differential
geometric case.
But we will describe $M[V]$ and the different strata explicitly as point
sets and present coordinate charts that give $M[V]$ a structure of
manifold with corners.  Our goal here will be to be explicit, rather
than to provide all details in proofs since an extension of the blowup
procedure in [\FM] to manifolds with corners gives a simple conceptual
proof.
To perform the anomaly calculation in \S6, we use
Stokes theorem; for this all we really nead are the coordinates on the
codimension $1$ open strata.

\newsubsec\SSDMV{Definition of $M[V]$ as a Closure}

For the remainder of this section, the integer $V$ will be fixed.
In accordance with our application to Feynman graphs,
elements of the set $\uV\equiv\{1,...,V\}$ will be referred to
as vertices.  The set $M^V$ is by definition
$M^{\uV}$, the set of maps from $\uV$ to $M$.
For $S$ a subset of $\uV$ containing at least two vertices,
$\Delta_S$ will denote
the smallest diagonal in $M^S=\Map(S,M)$ consisting of constant maps
from $S$ to $M$.  Similarly,
$\bar\Delta_S\subset M^\uV$ will denote the diagonal in $M^\uV$ which
maps to $\Delta_S$ under the projection map from $M^\uV$ to $M^S$.
$\bar\Delta_S$ consists of maps from $\uV$ to $M$ which send all vertices
in $S$ to the same point in $M$.

The blowup of $M^S$ along the diagonal $\Delta_S$ will be called
$\Bl(M^S,\Delta_S)$.
It has interior $M^S\setminus\Delta_S$ and boundary
$S(N(\Delta_S\subset M^S))$,
the sphere bundle of the normal bundle to the small diagonal in $M^S$.
This differential geometric blowup distinguishes a direction in
$N(\Delta_S)$ from its negative.  Let $\Bl_a(M^S,\Delta_S)$ denote the
algebraic geometric blowup used in [\FM].  There is a natural map
$\phi_S:\Bl(M^S,\Delta_S)\rightarrow \Bl_a(M^S,\Delta_S)$ which
identifies two rays in $N(\Delta_S)$ in opposite directions.

Since the projection map $\pi_S: M^\uV\rightarrow M^S$ maps $M^V_0$ to the
interior of $\Bl(M^S,\Delta_S)$ for $S\subset\uV$ with $|S|\ge 2$,
it determines a map
$ \pi_{0,S}: M^V_0 \rightarrow \Bl(M^S,\Delta_S)$.
Putting these maps together with the inclusion
$\vec f_0:M^V_0\rightarrow M^\uV$, we obtain an embedding
$$ M^V_0 \subset M^{\uV} \times \prod_{|S|\ge 2} \Bl(M^S,\Delta_S)
.\etag\embed$$
The space on the right hand side of \embed\ will be called $\cB$.
Since $\cB$ is a product of manifolds with boundary, it is a manifold
with corners.
$M[V]$ is defined to be the closure of the image of $M^V_0$ in $\cB$.
In the algebro-geometric context, there is a corresponding space
$$ \cB_a = M^{\uV} \times \prod_{|S|\ge 2} \Bl_a(M^S,\Delta_S)
\etag\abdef$$
and a continuous map $\phi:\cB\rightarrow\cB_a$.
The map $\phi$ sends $M[V]$ onto the Fulton--Macpherson compacitification
$M_a[V]$, the closure of $M_V^0$ in $\cB_a$.

In [\FM], $M_a[V]$ is shown to be equal to
a sequence of algebro-geometric blowups of $M^V$.
When $M$ is nonsingular, the blowups are on smooth submanifolds
and hence $M_a[V]$ is a smooth manifold, in fact a submanifold of $\cB_a$.
This procedure carries over to the differential geometric setup
using manifolds with corners, so that $M[V]$ is equal to a succession
of blowups of $M^V$ along submanifolds with corners and is a submanifold
with corneres of $\cB$.

\bigskip

We now describe $M[V]\subset\cB$ explicitly.
A point in $\cB$ is of course a pair $(\vec x, \{\vec x_{B,S}\})$,
where $\vec x$ is an element of  $M^\uV$ and
$\vec x_{B,S}$ is an element of $\Bl(M^S,\Delta)$
for each $S\subset\uV$ with $|S|\ge 2$.
Given such a pair,
let $\vec x_S$ be the image of $\vec x_{B,S}$ under the blowdown map
from $\Bl(M^S,\Delta_S)$ to $M^S$.  If $\vec x_S$ does not lie in
$\Delta_S$, $\vec x_{B,S}$ just equals $\vec x_S$.  Otherwise
$\vec x_{B,S}$ also contains the information of a point
in the fiber of $S(N(\Delta_S\subset M^S))$ at $\vec x_S$.

Given $\vec x_S\in\Delta_S$,
let $z\in M$ be the common location of all the vertices in $S$.
The fiber $N(\Delta_S\subset M^S)|_{\vec x_S}$ may
be identified with $[T_z M]^S/T_z M$, the quotient of the set of maps
from $S$ to $T_z M$ by overall translations.  The sphere bundle is
then the further quotient of the set of nonzero elements of the normal
bundle by the group $\IR_+$ of dilations.   Given a point
$\vec u_S\in [T_z M]^S$,
its orbit under the combined actions of $T_z M$ and $\IR_+$ will
be written $[\vec u_S]$.  So
$S(N(\Delta_S\subset M^S))|_{\vec x_S}$ is the set of
orbits $[\vec u_S]$ such that not all of the components of $\vec u_S$ are the
same.  In the terminology of [\FM], $[\vec u_S]$ is called
a {\it screen} for $S$ at $z$.  Given a metric on $M$, screens may be
uniquely represented by vectors $\vec u_S$ chosen to have
norm $1$ and to be orthogonal to $\Delta_S$.
It will be convenient to set $u_{S,j}=0\in TM_{x_j}$ when $j\notin S$, so
that now $\vec u_S\in T_{\vec x}M^\uV$ has norm $1$ and is orthogonal
to $\bar\Delta_S$.

\newsubsec\SSPMV{Description of $M[V]$ as a Point Set}

Which points in $\cB$ lie in $M[V]$?
Let $\cC$ be the subset of $\cB$
consisting of points $(\vec x,\{\vec x_{B,S}\})$
satisfiying the following two conditions.

\noindent C1: {\it $\vec x_S = \vec x|_S$
 for $S\subset\uV$, $|S|\ge 2$.
 }

\noindent{C2: \Underline{Compatibility condition for screens}.}
{\it
Suppose that $S'$ is a subset of $S$ with $|S'|>1$,
$\vec x$ maps all vertices in $S$ to $z$,
and the values of $\vec u_S$ on the vertices in $S'$ are not all equal.
Then
$[\vec u_{S'}]$ equals the restriction, $[\vec u_S|_{S'}]$, of
the screen for $S$ to a screen for $S'$.
 }

We now sketch an argument showing $M[V]=\cC$.
Since condition C1 holds for points in $M^V_0\subset\cB$, it holds
for $M[V]$.
Since $\vec x_S$ is determined by $\vec x$ for points in $M[V]$, we may
consider $M[V]$ to be a set of pairs
$(\vec x,\{[\vec u_S];\, \vec x|_S \in\Delta_S\})$.

\bigskip

Suppose $\vec x(t)$ is a smooth path in
$M^{\uV}$ parameterized by $t$ in $\IR_{\ge 0}$ (the non-negative reals)
with the property that
$\vec x(0)=(z,z,...,z)\in M^\uV$ and $\vec x(t)\in M^\uV_0$ for
$t>0$.
Choose local coordinates on $M$ with the origin centered about $z$.
The Taylor expansion of the components of $\vec x(t)$ about $t=0$
takes the form
$$ x_i(t) = v_i(1) t + v_i(2) t^2 + ..., \qquad\qquad\hbox{for $i\in\uV$}
.\etag\taylor$$
Although it is by a coordinate system dependent operation,
the components of $v_i(k)$ determine a vector in $T_z M$.
Let $n(S)$ be the smallest integer so that
$ v_i(n(S)) \ne v_j(n(S))$ for some $i,j\in S$.
Suppose now that the path $\vec x(t)$ is chosen so that $n(S)<\infty$
for all $S$, $|S|>1$.
Then the limit
$$ (\vec x_S,[\vec u_S]) \equiv \lim_{t\rightarrow 0^+} \pi_{0,S}(\vec x(t))
\etag\lambl$$
exists;
$\vec x_S$ maps every vertex in $S$ to $z$ and
$\vec u_S$ is the map that sends the vertex $i\in S$
to $v_i(n(S))$.

The hypothesis in the compatibility condition for screens which requires
that the values of $\vec u_S$ on the vertices in $S'$ are not all equal
is equivalent to the condition that $n(S)=n(S')$.
Hence $\{[\vec u_S]\}$ satisfies $C_2$.  So the limit in $M[V]$ of
$\vec x(t)$ as $t\rightarrow 0^+$, which equals
$(\vec x,\{[\vec u_S]; \vec x|_S\in\Delta_S\})$, lies in $\cC$.

Simple elaboration on this basic example, proves that all points in
$\cC$ can be obtained this way.
This shows $M[V]\supset\cC$.
We leave the reverse inclusion to the reader.
One needs to show that a limit point in $M[V]$ is the limit of a curve
$\vec x(t)$ as above using the compactness of the unit sphere
bundle in $T(M)$.

\newsubsec\SSSMV{Stratification of $M[V]$}

Having shown that $M[V]=\cC$, we can now decompose it into
a disjoint union of open strata,
$$ M[V] = \union_\cS M(\cS)^0
.\etag\stratif$$
Here $\cS$ is a collection of subsets of $\uV$, each subset containing
two or more elements, which are nested:
if sets $S_1, S_2$ belong
to $\cS$, they are either disjoint or else one contains the other.

The open strata $M(\cS)^0$ consists of the elements
$(\vec x, \{\vec x_{B,S}\})$ of $M[V]$ satisfying the following
conditions
\item{(i)} $\vec x|_S\in\Delta_S$ exactly when $S\subset S'$ for some
 $S'\in\cS$;
\item{(ii)}
When $S$ is the smallest set in $\cS$ containing $S'$,
$[\vec u_{S'}] = [\vec u_S|_{S'}]$.
\item{(iii)}
If $S_1, S_2\in\cS$ and $S_1\subset S_2$, then $\vec u_{S_2}|_{S_1}$
is a constant map.

Conditions (ii) and (iii) together say that the screens
$\{[\vec u_S]; S\in\cS\}$ are
independent and determine the remaining screens.

S1 and S2 below should now be clear; S3 and S4 follow from
our description of the manifold with corner structure on $M[V]$ in
the next subsection.

\noindent S1:
$M(\cS)^0$ is a smooth (noncompact) manifold
of codimension $|\cS|$ in $M[V]$, i.e. of dimension $3V-|\cS|$.

\noindent S2:
The closed strata $M(\cS)$, the closure of $M(\cS)^0$,
equals $\union_{\cT\supseteq\cS} M(\cT)^0$.

\noindent S3:
The codimension $k$
boundary to $M[V]$ is the union of the codimension $k$ closed strata,
$$ \del_k M[V] = \union_{\cS;\, |\cS|=k} M(\cS)
.\etag\clbdry$$

\noindent S4:
$ \del_k M[V]\setminus \del_{k+1} M[V]$ is the open set
in $\del_k M[V]$ given by $\union_{\cS;\, |\cS|=k} M(\cS)^0$.

For the codimension $1$ strata needed in the next section,
$\cS$ consists of a single set $S$ with $|S|>1$.  Then
$M(\cS)^0$ is the set of pairs
$(\vec x,\{[\vec u_S]\})$ for which
$x_i=x_j$ if and only if $i,j\in S$ and the components of
$\vec u_S$ are distinct and sum to zero.

\newsubsec\SSCMV{Coordinates on $M[V]$}

Let $c^{(0)}=(\vec x^{(0)},\{[\vec u_S^{(0)}]; S\in\cS\})$
be a point in $M[V]$ belonging to the open strata $M(\cS)^0$.
We now define coordinates on $M[V]$ in a neighborhood of
$c^{(0)}$.  The definition will make use of a choice of $g\in\Met$.
The collection of all coordinate systems as $c^{(0)}$ varies
defines the manifold with corner structure of $M[V]$.
This structure is independent of the choice of $g$.
Having fixed $g$, we may choose $\vec u_S^{(0)}$ to be the unique
representative of its screen with norm $1$ and satisfying
$\sum_{i\in S} u_{S,i}^{(0)} =0$.  Here $u_{S,i}^{(0)}$ is
the value of $\vec u_S^{(0)}$ at the point $i\in S$.

Define a map $\psi: M(\cS)^0\times [\IR_{\ge 0}]^{\cS}\rightarrow M^V$ by
$$\eqalign{
 \psi(c,\vec t) &= (\u{x}_1(c,t),..., \u{x}_V(c,t))
\cr
 \u{x}_i(c,t) &=\exp_{x_i}( \sum_{S\in\cS;\atop i\in S} \tilde t_S\, u_{S,i})
\cr
 \tilde t_S &= \prod_{S'\in\cS;\atop S'\supset S} t_{S'}
,}\etag\last$$
where $c=(\vec x,\{[\vec u_S]; S\in\cS\})$
and $\sum_{i\in S} u_{S,i}=0$, $||\vec u_S||=1$ for $S\in\cS$.

\medskip

\noindent\Bf{Lemma.}
{\it
There exists an open neighborhood $U$ of $c^{(0)}$ in $M(\cS)^0$ and
an open neighborhood $W$ of $\vec 0$ in $[\IR_{\ge 0}]^\cS$ so that the
restriction $\psi_0=\psi|_{U\times (W\setminus\del W)}$ maps into $M^V_0$
and is a diffeomorphism onto its image.
\endproof}

\noindent{\it Remark: }
It makes sense to claim that $\psi_0$ is a diffeomorphism since both
$M(\cS)^0$ and $\IR_{\ge 0}^\cS\setminus\del\IR_{\ge 0}^\cS$ are smooth
manifolds (without corners).

\noindent{\Underline{\it Proof:}}

By the inverse function theorem, it suffices to show that $U$ and $W$ may
be chosen so that
\item{(i)} $\psi_0$ maps into $M^V_0$.
\item{(ii)} The derivative of $\psi_0$ is injective.
\item{(iii)} $\psi_0$ is injective.

We need only consider the case when $\uV\in\cS$.
During the proof we will identify a screen $[\vec u_S]$
at a point $\vec x=(x,...,x)$ in the total diagonal in $M^V$
with its preferred representative $\vec u_S$ of norm $1$ satisfying
$\sum_{i\in S} u_{S,i}=0$.
Recall that we set $u_{S,i}=0$ for $i\notin S$ so that we may
view $\vec u_S$ as an element of  $TM^V_{\vec x}$.
Let $<\cdot,\cdot>$ denote the inner product on $TM^V$.

\noindent{\it Proof of (i).}
By the tubular neighborhood theorem, it suffices to show that,
for suitably small $U$ and $W$,
$$ \sum_{S\in\cS}\tilde t_S\, u_{S,i} \ne \sum_{S\in\cS}\tilde t_S\, u_{S,j}
\etag\difs$$
for $i\ne j$, $(\vec x,\{\vec u_S\})\in U$, and $\vec t\in W$.
Let $S_0$ be the smallest set in $\cS$ containing $i$ and $j$.
Since $u_{S,i}-u_{S,j}=0$ for $S$ not a subset of $S_0$,
the difference of the two sides of \difs\ is
$$\tilde t_{S_0}(u_{S_0,i}-u_{S_0,j})
  +\sum_{S\subsetneqq S_0}\tilde t_S(u_{S,i}-u_{S,j})
.\etag\omto$$
Note that for $S\subsetneqq S_0$, $\tilde t_S$ equals
$t_{S_0}\tilde t_S$ times a product of some other $t$'s.
Also $u_{S_0,i} - u_{S_0,j}\ne 0$.
Hence, one can choose $U$ and $W$ small enough so that
$|\tilde t_S(u_{S,i}-u_{S,j})|$ is much smaller than
$|\tilde t_{S_0} (u_{S_0,i}- u_{S_0,j})|$,
and therefore \omto\ is non-zero.

\noindent{\it Proof of (ii).}
Using the tubular neighborhood theorem again as well as the fact that the
map $\vec t\mapsto\vec{\tilde t}$ from
$\IR_+^\cS$ to $\IR_+^\cS$ is a diffeomorphism,
it suffices to show that the map
$(\{\vec u_S\},\vec{\tilde t})\mapsto
\sum_{S\in\cS}\tilde t_S u_S$ is injective at the tangent space level.
The derivative of this map in the direction of
$(\{\delta\vec u_S\},\delta\vec{\tilde t})$ is
$$\sum_{S\in\cS} \vec u_S\delta\tilde t_S +\tilde t_S\delta \vec u_S
.\etag\mbnzero$$
The fact that $\{\vec u_S\}$ is an orthonormal set of vectors in $TM^V$
implies that
$<\vec u_S,\delta\vec u_{S'}> =0$ for any $S,S'\in\cS$, and that
$<\delta\vec u_S,\delta\vec u_{S'}>=0$ for $S\ne S'$.
Hence, all the individual terms in \mbnzero\ are orthogonal.
Therefore \mbnzero\ is zero only if $\delta\vec{\tilde t}$ and
the $\delta\vec u_S$ all vanish.

\noindent{\it Proof of (iii).}
Using the tubular neighborhood theorem one more time, it suffices to show that
if $U$ and $W$ are suitably small and if
$((\vec x,\{\vec u_S\}),\vec t)$ and $((\vec x,\{\vec U_S\}),\vec T)$
are points in $U\times (W\setminus\del W)$ projecting to the same point
$\vec x\in TM^V$, then
$\sum_{S\in\cS}\tilde t_S\, u_S$ equals $\sum_{S\in\cS} \tilde T_S\, U_S$
only when $\tilde t_S =\tilde T_S$ and $u_S= U_S$ for all $S\in\cS$.
This follows because $u_S$ and $U_S$ have norm $1$ and
$<u_S,U_{S'}>=0$ for $S\ne S'$.
\endproof

\bigskip

\noindent\Bf{Theorem.}
{\it
The map $\psi_0$ of the previous lemma extends continuosly to a map
$\psi_B: U\times W\mapsto \Im(\psi_B) \subset M[V]$ so that
\item{T1.}
$\psi_B(c,\vec t) \in M(\cS')^0$ where $\cS'\equiv\{S\in\cS; t_S=0\}$.
\item{T2.} $\psi_B(c,\vec 0)=c$.
\item{T3.} $\psi_B$ is a homeomorphism.
\item{T4.} The set of maps $\psi_B$ as $c$ varies over $M[V]$ are
a system of coordinates on $M[V]$ giving it a structure of a manifold
with corners.
\item{T5.} The manifold with corner structure on $M[V]$ is indepent of
the choice of metric $g$.
\item{T6.} $M(\cS)^0$ is an open subset of the smooth part of the codimension
$|\cS|$ boundary of $M[V]$.
\item{T7.} The inclusion map of $M[V]$ in $\cB$ is smooth.
\endproof}

\noindent{\Underline{\it Outline of Proof:}}

Choose $(c,\vec t)\in U\times W$, $c=(\vec x,\{[\vec u_S]; S\in\cS\})$
and let $\cS'=\{S\in\cS'; t_S=0\}$ as above.

Let
$\vec\tau:[0,\infty)\rightarrow M^V$ be the smooth curve
$$\vec\tau(\epsilon)=\psi(c,\vec t_\epsilon)
,\etag\last$$
where $\vec t_\epsilon\in\IR_{\ge 0}^\cS$ is given by
$$(t_\epsilon)_S
 =\cases{ t_s &for $S\notin\cS'$
\cr
          \epsilon &for $S\in\cS'$
   }
.\etag\last$$
Let $\u{\vec x}=\vec\tau(0)=\psi(c,\vec t)$.
Observe that, when $\epsilon$ is a small positive number,
$\vec\tau(\epsilon)$ equals $\psi_0(c,\vec t_\epsilon)\in M^V_0$.
Therefore, if $\psi_B$ exists, it must equal
\eqna\pbmb
$$\eqalignno{
\psi_B(c,\vec t) &= (\u{\vec x}, \{[\u{\vec u}_S];
(\u{\vec x})|_S\in\Delta_S\})
&\pbmb{.1}\cr
\bnoalign{\hbox{where}}
(\u{\vec x}|_S,[\u{\vec u}_S])
  &=\lim_{\epsilon\rightarrow 0^+}\pi_{0,S}(\vec\tau(\epsilon))
\quad\hbox{for $S\subset\uV$, $|S|\ge 2$, $(\u{\vec x})|_S\in\Delta_S$.}
&\pbmb{.2}
}$$

The limit in \pbmb{.2}\ can be calculated
in terms of the Taylor series of $\tau(\epsilon)$
in the manner introduced in \S\SSPMV.
Write $\tau_i(\epsilon)=\exp_{x_i}(w_i(\epsilon))$, where
$$\eqalign{
 w_i(\epsilon) &=
  \sum_{S\in\cS;\atop i\in S}\epsilon^{m(S)}\,\tilde t'_S\, u_{S,i}
\cr
 m(S) &= |\{S'\in\cS; S'\supset S, t_{S'}=0\}| =|\{S'\in\cS'; S'\supset S\}|
\cr
 \tilde t'_S &= \prod_{S'\in\cS, t_{S'}\ne 0;\atop S'\supset S } t_{S'}
.}\etag\last$$

Let $v_i(n)$ be the coefficient of $\epsilon^n$ in $w_i(\epsilon)$.  Then
\eqna\vsws
$$\eqalignno{
 w_i(\epsilon) &= \sum_{n=0}^{|\cS|} v_i(n)\epsilon^n
&\vsws{.1}\cr
 v_i(n) &= \sum_{S'\in\cS; i\in S';\atop m(S')=n} \tilde t'_{S'}\, u_{S',i}
&\vsws{.2}
.}$$

Observe that $\u x_i=\exp_{x_i}(v_i(0))$.
Also observe that when $n=0$ the terms in
the sum \vsws{.2}\ have $m(S')=0$; so $\tilde t'_{S'}=\tilde t_{S'}$.
A little thought suffices to verify that
$$ \left[ {\u x}_i = {\u x}_j\right]  \Leftrightarrow
    \left[ x_i =x_j\hbox{ and } v_i(0)=v_j(0)\right] \Leftrightarrow
   \left[ i,j\in S' \hbox{ for some } S'\in\cS'\right]
.$$
So the limit in \pbmb{.2}\ must be calculated for $S\subset\uV$ with
$|S|\ge 2$ and $S\subset S_0$ for some $S_0\in\cS'$.
Fix such an $S$ until further notice.

Let $z=\u x_i$ for $i\in S$ and $F_z:TM_{x_i}\rightarrow TM_z$ be
the vector space isomorphism
$$ F_z(w) =\tot{\kappa}\bigg|_{\kappa=0}\exp_{x_i}(v_i(0)+\kappa w)
\qquad\hbox{for $w\in TM_{x_i}$}
.\etag\last$$
(I.e. $F_z(w)$ is obtained from $w$ by transport using the Jacobi equation,
the geodesic deviation equation.)
Then
\eqna\ttau
$$\eqalignno{
\tau_i(\epsilon) &= G\left(\sum_{n=1}^{|\cS|} F_z(v_i(n))\epsilon^n\right),
&\ttau{.1}\cr
\bnoalign{\hbox{where}}
G(a) &=\exp_{x_i}(v_i(0) + (F_z)^{-1}(a))\hbox{ for $a\in TM_z$.}
&\ttau{.2}
}$$
The argument of $G$ in \ttau{.1}\ is the version in the present context
of the right hand side of \taylor.  Using the map $G$ simply provides an
invariant way of identifying points near $z$ with points in $TM_z$.
If we choose $g$ to be flat near $x_i$ and work in flat coordinates,
$G$ and $F_z$ become trivial.

Set $ n(S)=\min\{n; v_i(n)\ne v_j(n)\hbox{ for some }i,j\in S\}$
as in the paragraph below \taylor.
Using the facts that
$$S'\subset S\Rightarrow m(S')\ge m(S)
$$
and
$$
[ S'\in\cS,\, i,j\in S\cap S',\, m(S')<m(S) ]
\Rightarrow u_{S',i} = u_{S',j}
$$
and the technique used to prove point (iii) in the previous lemma,
it's not hard to show that $n(S)=m(S)$.
Hence, as in the sentence after \lambl,
$$\u{\vec u}_{S,i}= F_z(v_i(n(S)))
  =\sum_{S'\in\cS; i\in S'\atop m(S')=m(S)}\tilde t'_{S'} F_z(u_{S',i})
\etag\finu$$
for $i\in S$.
Note that the sets $S'$ in the sum in \finu\ need not necessarily be
contained in or contain $S$.

Specializing \finu\ to the case when $\vec t=0$, we obtain
$\u{\vec u}_{S}={\vec u}_{S_1}|_{S}$, where $S_1$ is the smallest element of
$\cS$ containing $S$.  This verifies T2.

Using \finu we find:  if $S_1, S_2\subset S_0\in\cS'$,
$|S_1|\ge 2$ and $S_2\subsetneqq S_1$ then
$$
\left[\vec u_{S_1}|_{S_2}\hbox{ is constant}\right]\Leftrightarrow
\left[ \exists S_3\in\cS'; S_2\subset S_3\subsetneqq S_1 \right]
.$$
This statement is equivalent to T1.

The verification of T3, that the extension $\psi_B$ of $\psi_0$
defined by \pbmb{.1}\ and \finu\ is a homeomorphism, is
an exercise in point set topology.

To prove T4 and T5 it is necessary to show that the overlap map
between coordinate charts $\psi_B, \psi'_B$, associated to choices
of $c,c'\in M[V]$ and $g,g'\in Met$, is a diffeomorphism from one
manifold with corners (an open subset of $U\times W$) to another
(an open subset of $U'\times W'$).  We will not carry out this very
tedious exercise here.  It can be derived more conceptually by using
the map $\psi_{B,a}=\phi\circ\psi_B:U\times W\rightarrow M_a[V]$
which is given explicitly by \finu\ together with
$\psi_{B,a}(c,\vec t) = (\u{\vec x}, \{[\u{\vec u}_S]_a \})$, where
$[\u{\vec u}_S]_a$ is the orbit of $\u{\vec u}_S$ under the combined
action of translation by $T_z M$ and multiplication by
$\IR\setminus\{0\}$ (rather than $\IR_+$).
One can check that $\psi_{B,a}$ extends to a coordinate chart for
$M_a[V]$ (by allowing the $t_S$'s to be negative).  The overlap maps
of the $\psi_{B}$'s are the restriction to non-negative $t_S$'s of
the overlap maps for the $\psi_{B,a}$'s and hence are smooth.

Finally, T6 and T7 follow by inspection.
\endproof

\bigskip
\newsec\Stheorem{Main Theorem}
\medskip

Our first basic result in [\AS] was that the integrals defining
$I(\G)$ are convergent despite the singularities near the
union of all the diagonals in $M^V$.
In fact, we proved a stronger version of this in [\AS]
using power counting techniques of physics.  We showed that the
integral
$\int_{M^V} \Tr^{(v)}(\cI(\LG)\Psi)$ converges for any
smooth $\Psi\in\Omega^*(M^V,A_V^*)$ and any $\G$
(not necessarily trivalent).  In the language of
quantum field theory, this says that Chern-Simons perturbation theory
is finite.

The main result of this paper is to prove the conjecture made in [\AS]
that the dependence of $I_l^{conn}$ on the arbitrary
choice of $g$ could be cancelled by subtracting a local counterterm which
is an appropriate multiple of the ``gravitational'' Chern--Simons invariant
$\cs_{grav}(g,s)$ of the metric connection on $M$, defined using a
homotopy framing $s$ of $TM$ (see [\AS]).  Stated another way,
we have our main theorem.
\medskip

\noindent\Bf{Theorem.}
{\it
There is a constant $\beta_l$ depending only on $l$ and
the bi-invariant inner product $<\,,\,>_{\Lie(G)}$ on $\Lie(G)$ so that
the quantity
$$\hat I_l^{conn}(M, A_0 ,s) \equiv
  I_l^{conn}(M, A_0,g) - \beta_l\cs_{grav}(g,s)
$$
is independent of $g$.  $\hat I_l^{conn}$ is therefore a topological invariant
depending on the choice of
manifold $M$, homotopy framing $s$, and flat connection $A_0$.
}

{\it
When $l$ is odd, $\beta_l=0$.
\endproof}

\noindent{\it Remark: }
The naturality
of our construction implies that the values of $\hat I_l^{conn}$ agree for two
choices of $(M, A_0, s)$ which are related by a principal bundle automophism.
(Although the automorphisms must be differentiable, we use the term
topological invariant since it is more standard in this context.)

\medskip

The proof of the theorem has three steps.
First, we shall rewrite $I_l$ as a push-forward
by integration over $M[V]$ of a
closed form on $M[V]\times\Met$ constructed from $\tilde L$.
Next we shall apply Stokes theorem to write the anomaly $d_{\Met} I_l$
as an integral over the boundary of $M[V]$.
Then we shall use the explicit descriptions
of the propagator singularities (i.e. $\tilde L|_{\del B_2\times\Met}$)
and of $\del M[V]$ to calculate $d_{\Met} I_l$.
This result will imply that
$$ d_{\Met} I_l^{conn} = \beta_l \cs_{grav}(g,s)
$$
as desired.

\medskip
\subsubsec{Step 1: Rewriting $I_l$}

Shortly we will define the total propagator $\tilde L_{C,tot}$
on the compactification $M[V]$.
It belongs to $\Omega^*(M[V],\tilde A_V^*)$.
$\tilde A_V^*$ stands for the bundle $A_V^*$ pulled back to
either $M^V\times\Met$ or, in this case, to $M[V]\times\Met$.
$\tilde L_{C,tot}$ is characterized by the facts that it is
smooth on all of $M[V]$ and that it agrees with
$\tilde L_{tot}$ (the analog of $L_{tot}$ defined using the extended
propagator)
on $M^V_0\times\Met$.

Having defined $M[V]$ and $\tilde L_{C,tot}$, we may rewrite $I_l$ in terms
of them:
$$ I_l = \int_{M[V]} \Tr^{(V)}(\tilde L_{C,tot}{}^I)
.\etag\rewrite$$
The operator $\Tr^{(V)}$ is the same as the operator
$\Tr^{(V)}$ defined previously but now maps
$\Omega^*(M[V], \tilde A_V^*)$ to $\Omega^*(M[V])$.

The integral in \rewrite\ agrees with
$\int_{M^V}\Tr^{(V)}(L_{tot}{}^I)$.  Since the integrand has
degree $3V=\dim(M^V)$ as a differential form,
the integral picks out the piece of
$\tilde L_{tot}{}^I$ of degree $0$ in the $\Met$ directions.
This is precisely $L_{tot}{}^I$.  Thus \rewrite\ agrees with the
previous definition of $I_l$.

Now we define $\tilde L_{C,tot}$.
As one would expect, it is a double sum
$$\tilde L_{C,tot} = \sum_{i,j=1}^V \tilde L_{C,\{i,j\}}
\etag\tlc$$
of pieces $\tilde L_{C,\{i,j\}}\in\Omega^2(M[V]\times\Met, A_V^2)$.
$\tilde L_{C,\{i,j\}}$ smoothly extends
$\tilde L_{ab}(x_i, x_j) j^a_{(i)} j^b_{(j)}|_{M^V_0}$ to all of $M[V]$,
as follows.

Let $\pi_{B,\{i,j\}}: M[V]\rightarrow \Bl(M^{\{i,j\}},\Delta_{\{i,j\}})$
be the map
$$\pi_{B,\{i,j\}}((\vec x,\{\vec x_{B,S};S\subset\uV,|S|>2\}))
  =\vec x_{B,\{i,j\}}
\etag\last$$
and $f_{B,\{i\}}:M[V]\rightarrow M^{\{i\}}$ be the map
$$f_{B,\{i\}}((\vec x,\{\vec x_{B,S};S\subset\uV,|S|>2\}))
  =x_i
.\etag\last$$
The trivial cross product of these maps with $\Met$ will be denoted
by $\tilde\pi_{B,\{i,j\}}$ and $\tilde f_{B,\{i\}}$.

For $i\ne j$, $\tilde L_{C,\{i,j\}}$ is given by
$$ \tilde L_{C,\{i,j\}}=
(\pi_{B,\{i,j\}})^* (\tilde L_{Bs,\{i,j\}})
 =(\tilde L_B)_{ab}(\vec x_{B,\{i,j\}}) j^a_{(i)} j^b_{(j)}
.\etag\lctwo$$
Here $\tilde L_{Bs,\{i,j\}}$ is a copy of
the ``extended super-propagator'' $\tilde L_{Bs}=s(\tilde L_B)$,
but for the vertices $i,j$ rather than $1,2$.
For $i=j$, the appropriate definition is
$$\tilde L_{C,\{i,i\}}
  =(f_{B,\{i\}})^* (\tilde\rho_{s,\{i\}})
  =\tilde\rho_{ab}(x_i,x_i) j^a_{(i)} j^b_{(i)}
.\etag\alc$$
$\tilde\rho_{s,\{i\}}$ here is a copy of $\tilde\rho_s=s(\tilde\rho)$
belonging to
$\Omega^*(M^{\{i\}}\times\Met,\Lambda^2(\tilde\adp_1\otimes\tilde\adp_2))$.

For notational convenience in \lctwo, \alc, and below, we have not written the
argument $g$ explicitly.

\subsubsec{Step 2: Stokes Theorem}

Let $d_\Met$ and $d_{M[V]}$ be the exterior derivative operators.
Since $\tilde L_{C,tot}$ is
covariantly closed, the integrand in \rewrite\ is closed.  Hence,
$$\eqalign{
  d_{\Met} I_l &= c_{V,I}\int_{M[V]} d_{\Met}\Tr^{(V)}(\tilde L_{C,tot}{}^I)
\cr
	       &= c_{V,I}\int_{M[V]} -d_{M[V]}\Tr^{(V)}(\tilde L_{C,tot}{}^I)\cr
	       &= - c_{V,I}\int_{\del M[V]} \Tr^{(V)}(\tilde L_{C,tot}{}^I)
.}\etag\aone$$

\subsubsec{Step 3: Calculation of the Anomaly}

Because $\tilde L_{C,tot}$ is smooth, we are free to replace $\del M[V]$
in \aone\ by the open dense subset $\del M[V]\setminus \del_2 M[V]$.
The latter is the disjoint union of the codimension one open strata:
$$ \del M[V]\setminus \del_2 M[V]
  = \union_{\uV'\subset \uV,\; |\uV'|\ge 2} M(\{\uV'\})^0
  \subset_{{open\atop dense}}\del M[V]
.\etag\last$$
Furthermore, two different choices of
$\uV'$ which differ only by a permutation of $\uV$ give equal contributions.
Therefore, by including a combinatorial factor, we may restrict to the
standard choices $\uV'=\{1,...,V'\}$, ($2\le V' \le V$).
Thus, we obtain
$$ d_{\Met} I_l = -c_{V,I}\sum_{V'=2}^V {\textstyle {V\choose V'}}
    \int_{M({\uV'})^0} \Tr^{(V)}(\tilde L_{C,tot}{}^I)
.\etag\atwo$$

The term in \atwo\ with a given $V'$
is the contribution to the anomaly from the regions where $V'$ points
coincide.  It will be useful to introduce names
$V''\equiv V-V'$ and $\uV''\equiv \{V'+1,...,V\}$
for the number of and the label set for the points not coinciding.

Recall from \S\SBV\ that $M(\{\uV'\})^0$ equals the set of
$(\vec x,\{[\vec u_{\uV'}]\})$ where:
\item{(i)}
$\vec x=(x_1,...,x_V)$ is an element of $M^V$ with
$x_1$ through $x_{V'}$ all equal to some $z$ in $M_z$
(which is just a disjoint copy of $M$ labeled by $z$) and
all pairs $x_i$, $x_j$ distinct otherwise;
and
\item{(ii)}
$[\vec u_{\uV'}]$ is an element of the fiber of the sphere bundle
$S([TM_z]^{\uV'}/TM_z)$ at $z$
represented by a vector
$(u_1,...,u_{V'})\in [T_zM_z]^{\uV'}$ with no two components equal.

\noindent For $i\ne j$, we also have
$$ \vec x_{B,\{i,j\}} =\pi_{B,\{i,j\}}(\vec x,\{[\vec u]\}) =
  \cases{
 ((z,z),[(u_i,u_j)])\,\in\del Bl(M^{\{i,j\}},\Delta_{\{i,j\}})
&$i,j\in\uV'$ \cr\casespace
    (x_i,x_j)\,\in Bl(M^{\{i,j\}},\Delta_{\{i,j\}}) \setminus
      			\del Bl(M^{\{i,j\}},\Delta_{\{i,j\}})
&\hbox{otherwise.}
    }
\etag\last$$
As a particular case of the bottom line,
$\pi_{B,\{i,j\}}(\vec x,\{[\vec u_{\uV'}]\}) = (z,x_j)$
for $i\le V' < j$; and similary for $i$ and $j$ reversed.

This description of $\pi_{B,\{i,j\}}$ on
$M(\{\uV'\})^0$ allows us to write
$$\tilde L_{C,\{i,j\}}=\cases{
\tilde\rho(x_i,x_i) j^a_{(i)}\, j^b_{(i)}	 	 &$i=j>V'$
\cr\casespace
\tilde\rho(z,z) j^a_{(i)}\, j^b_{(i)}			 &$i=j\le V'$
\cr\casespace
\tilde L_{ab}(x_i,x_j)\, j^a_{(i)}\, j^b_{(j)}		 &$i,j>V'$, $i\ne j$
\cr\casespace
\tilde L_{ab}(z,x_j)\, j^a_{(i)}\, j^b_{(j)}		 &$i\le V'$, $j>V'$
\cr\casespace
\tilde L_{ab}(x_i,z)\, j^a_{(i)}\, j^b_{(j)}		 &$i > V'$, $j\le V'$
\cr\casespace
[\tilde\lambda(z,[(u_i,u_j)] )\delta_{ab}
 +\tilde\rho_{ab}(z,z)]\, j^a_{(i)}\, j^b_{(j)}     	 &$i,j\le V'$, $i\ne j$.
  }
\etag\last$$

Thus we may decompose $\tilde L_{C,tot}$ into terms coming from the explicit
propagator singularity and remaining ``regular'' terms,
\eqna\expandl
$$\eqalignno{
  \tilde L_{C,tot} =& \tilde L_{sing, V'} + \tilde L_{reg, V''}
&\expandl{.1}\cr
  \tilde L_{sing,V'} =& \sum_{{i,j\le V' \atop i\ne j}}
	\tilde\lambda(z,[(u_i,u_j)])\, [j^a_{(i)}\wedge j^a_{(j)}]
  &\expandl{.2}\cr
  \tilde L_{reg,V''}
  =& \tilde\rho_{ab}(z,z)\, J^a\wedge J^b
     +\sum_{i> V'} \tilde\rho_{ab}(x_i,x_i) j^a_{(i)} \wedge j^b_{(i)}
&\expandl{.3}\cr
   & +\sum_{j> V'} \tilde L_{ab}(z,x_j)\, J^a\wedge j^b_{(j)}
     +\sum_{i> V'} \tilde L_{ab}(x_i, z)\, j^a_{(i)} \wedge J^b
\cr
   & +\sum_{{i,j> V' \atop i\ne j}} \tilde L_{ab}(x_i,x_j)\,
      j^a_{(i)}\wedge j^b_{(j)}
,\cr
\bnoalign{\hbox{where}}
J^a =& \sum_{1\le i\le V'} j^a_{(i)}
.&\expandl{.4}
}$$

\noindent The important properties of \expandl{}\ which we need are the
following.
\item{P1.}
$\tilde L_{reg,V''}$ only depends on $(z, J)$ and the $(x_i, j_{(i)})$
 for $i> V'$.
\item{P2.}
$\tilde L_{sing,V'}$ only depends on the $(x_i,j_{(i)})$ for $i\le V'$,
and on $[\vec u_{\uV'}]$.
\item{P3.}
Each term in the sum \expandl{.2}\ defining $\tilde L_{sing,V'}$ factors
into a ``group theory piece'' ($j^a_{(i)}\wedge j^a_{(j)}$) times a
``manifold piece'' ($\tilde\lambda(z,[u_i,u_j])$).
\item{P4.}
$\tilde L_{sing,V''}$ is invariant under diagonal gauge transformations,
that is gauge transformations that acts the same on all factors
$\tilde\adp_1, ..., \tilde\adp_{V'}$.

\noindent P4 follows from the invariance of the Lie algebra metric.

Plugging the first line of \expandl{}\ into \atwo\ and expanding using
the binomial theorem, one finds
$$ d_{\Met} I_l = -\sum_{V'=2}^V \sum_{I'=0}^I c_{V',I'}\, c_{V'',I''}
    \int_{S(TM_z{}^{\uV'} /TM_z)\times M^{\uV''}} \Tr^{(V)}
\left( [\tilde L_{sing,V'}]^{I'}\wedge [\tilde L_{reg, V''}]^{I''} \right)
,\etag\athree$$
where $I''=I-I'$.
The domain of integration indicated gives the same result as
$M({\uV'})^0$ which is an open dense subset.

The next step consists of breaking the integral up into three parts:
(1) an integral over $(x_{V'+1},...,x_V)$ in $M^{\uV''}$,
together with the Lie algebra traces for $j_{(V'+1)}, ... , j_{(V)}$
in $\tilde\adp^{\uV''}=\tilde\adp_{V'+1}\oplus ...\oplus \tilde\adp_V$;
(2) an integral over $[\vec u]$ in $S(TM_z{}^{\uV'}/TM_z)|_z$
for fixed $z$, together with the contractions over the
nondiagonal directions in
$\tilde\adp^{\uV'}=\tilde\adp_1\oplus ...\oplus\tilde\adp_{V'}$;
and
finally (3) an integral over $z$ in $M_z$  together with contractions for
$J$ which belongs to the diagonal directions
$\tilde\adp_{J}\subset \tilde\adp^{\uV'}$.

Before proceeding we explain the phrase ``contraction over the diagonal
directions''.  Write $\tilde\adp^{\uV''}={\Bf{h}}\oplus\tilde\adp_J$,
where
$${\Bf{h}}=\{(j_1,...,j_{\uV'})\in\tilde\adp^{\uV'};
    \, \sum_{i=0}^{V'} j_{(i)} = 0\}
\etag\last$$
and $\tilde\adp_J$ is its orthogonal compliment.  The subspace
$\tilde\adp_J$ is the space of diagonal directions, that is
$$\tilde\adp_J =
\{(j_{(1)},...,j_{(V')})\in \tilde\adp^{V'}; j_{(r)} = j_{(s)},\, r,s\in \uV'\}
.\etag\last$$
So
$$\Lambda^{2I'}(\tilde\adp^{\uV'}) =
 \sum_r \Lambda^{2I'-r}({\Bf{h}}) \otimes \Lambda^r(\tilde\adp_J)
.\etag\last$$
Contraction over the nondiagonal directions means interpreting
$\eta\in\Lambda^{2I'}(\tilde\adp^{\uV'})$ as a linear function
acting on $\omega\in\Lambda^E(\tilde\adp_J)$ by wedging to get
$\eta\wedge\omega\in \Lambda^{2I'+E}(\tilde\adp^{\uV'})$
and applying $\Tr^{(V')}$, giving a real number
$\Tr^{(V')}(\eta\wedge\omega)$ (which vanishes unless
$E=3V'-2I'$).

We may write
$$ d_{\Met} I_l = \sum_{V'=2}^V \sum_{I'=0}^I
   \int_M < \bar A_{V',I'}\> , \>\bar C_{V'', I''}>
.\etag\afour$$
We now discuss the two pieces $\bar A_{V',I'}$ and
$\bar C_{V'',I''}$ of this equation.

$\bar C_{V'',I''}$ is the result of pushing forward
$[\tilde L_{reg,V''}]^{I''}$, considered (by P1) as an element of
$\Omega^{2 I''}(M_z\times M^{\uV''}\times\Met,
		\Lambda^{2 I''}(\tilde\adp_J\times\tilde\adp^{\uV''})) $,
by integration over $M^{\uV''}$ and contraction on $\tilde\adp^{\uV''}$,\foot{
  In physical parlance, $\bar C_{V'',I''}$ is the
  untruncated Green's function with $E$ external legs
  and at order $(I''-E)-V''+1$ in perturbation theory,
  evaluated on the superspace diagonal (meaning that all external vertices and
  generalized polarization tensors agree).
  }
$$\bar C_{V'',I''}(z,J)
 =c_{V'',I''} \int_{(x_{V'+1},...,x_V)\in M^{V''}}
  \Tr_{V'+1}\circ ...\circ\Tr_V \left([\tilde L_{reg,V''}]^{I''}\right)
.\etag\defC$$
Since the integration subtracts manifold form degree $3V''$, and the
Lie algebra traces subtract Lie algebra form degree $3V''$,
$\bar C_{V'',I''}$ belongs to
$\Omega^E(M_z\times\Met,\Lambda^E(\tilde\adp_J))$,
where
$$E = 2 I'' - 3 V'' = 3V' - 2I'
.\etag\defE$$

Similarly $\bar A_{V',I'}$ is the push forward of $\tilde L_{sing,V'}{}^{I'}$,
considered (by P2) as an element of
$\Omega^{2I'}(S(TM_z{}^{\uV'}/TM_z)\times\Met,
 \Lambda^{2I'}(\tilde\adp^{\uV'}))$,
by integration over the fibers of $S(TM_z{}^{\uV'}/TM_z)\rightarrow M_z$
and contractions
in the non-diagonal directions, as explained above.
Thus,  for
$\omega(z,J)\in\Lambda^*(\tilde\adp_J)|_z$, we have
$$<\bar A_{V',I'}(z,J),\omega(z,J)>
 = - c_{V',I'} \int_{[\vec u]\in S(TM_z{}^{\uV'}/TM_z)}
   \Tr_1\circ ...\circ \Tr_{V'}
   \left([\tilde L_{sing,V'}]^{I'}\wedge\omega(z,J)\right)
.\etag\defA$$
The degree as a differential form of $\bar A_{V',I'}$ is
$$ 2I' - \dim(S(TM_z{}^{\uV'}/TM_z)|_z)
    = 2 I' - (3V'-4) = 4-E
,\etag\lat$$
whereas it pairs with Lie algebra forms of degree $3V'-2I'=E$.
$\bar A_{V',I'}$ is invariant under gauge transformations,
as follows from invariance of $\tilde L_{sing,V'}$ (see P4) and
the operators $\Tr_i$ for $i\le V'$.
Putting this together,
$\bar A_{V',I'}$ belongs to
$\Omega^{4-E}(M_z\times\Met,[\Lambda^E(\tilde\adp_J)^\dual]^{invt})$,
where $W^\dual$ denotes the dual of a vector space $W$.

\medskip

One can now expand each factor of $\tilde L_{sing,V'}$ appearing in
\defA\ using \expandl{.2}\ and rewrite the resulting sum as a sum
over labeled graphs, as was done in arriving at \graphint.
By property P3 above, the contribution of each graph factors
into a manifold piece times a group theory piece.
The result is
\eqna\Ais
$$\eqalignno{
\bar A_{V',I'} &= - c_{V',I'} \sum_{\LG'} A_{mfld}(\G') A_{gp}(\G'),
   \hbox{ where}
&\Ais{.1}\cr
A_{mfld}(\G') &\equiv
 \int_{\{[\vec u]\}}\prod_{e=1}^{I'}\tilde\lambda(z,[u_{i_e},u_{j_e}])
  \in \Omega^{4-E}(M_z\times\Met),
&\Ais{.2}\cr}$$
and
$$
A_{gp}(\G') \equiv \Tr_1\circ ...\circ Tr_{V'}\circ
	\prod_{e=1}^{I'} j^a_{(i_e)} j^a_{(i_e)}
  \in\left[\Lambda^E(\tilde\adp_J)^\dual\right]^{invt}
   \cong \Lambda^E(\Lie(G)^\dual)^{invt}.
\eqno\Ais{.3}$$
The sum in \Ais{.1}\ is over all labeled graphs $\LG'$ with $V'$ vertices
and $I'$ edges which have no vertices of valency greater than $3$
(and also no edges connecting a vertex to itself).
$E$ is the number of external edges the graphs have, i.e. the
number of edge ends that need to be attached to any of the $\G'$ to make it
a trivalent graph.  These graphs have
$$ l' = I' - V' +1
\etag\last$$
loops.
Note that
$$ V'= 2(l'-1)+E,\,\hbox{ and}\qquad I'=3(l'-1)+E
.\etag\last$$
Also note that the graphs may be assumed connected since the integral
\Ais{.2}\ vanishes for $\G'$ disconnected.  This vanishing follows because
the integrand is annihilated by
interior product by the non-trivial vector field which scales
the $u_i$ for $i$ labeling one of the vertices in a connected component
of $\G'$ (see [\AS]).

The expression on the right of \Ais{.3}\ is an operator that when acting
on $\Lambda^E(\tilde\adp_J)$ produces a number, as in \defA.
Since it is gauge invariant and its explicit form does not depend on $z$,
it may be viewed as an element of the fixed space
$[\Lambda^E(\tilde\adp_J)^\dual]^{invt}\cong\Lambda^E(\Lie(G)^\dual)^{invt}$.
$A_{gp}(\G')$ only depends on the (unlabeled) graph $\G'$,
the group $G$, and the invariant metric $<\,,\,>_{\Lie(G)}$
 on $\Lie(G)$.  Its value remains unchanged if we
replace $G$ by its semisimple part $G_{ss}$ (and $<\,,\,>_{\Lie(G)}$
by its restriction
to $G_{ss}$).  This follows because the structure constants
(appearing in $\Tr_{i}$) in the direction
of the $U(1)$ factors of $G$ vanish.  With this replacement of $G$
by $G_{ss}$, $A_{gp}(\G')$ becomes an element of
$\Lambda^E(\Lie(G_{ss})^\dual)^{invt}$, considered as a subspace of
$\Lambda^E(\Lie(G)^\dual)^{invt}$.

Note that $A_{mfld}(\G')$ is a characteristic polynomial of $\tilde\BOmega$.
In other words,
$$ A_{mfld}(\G') = P_{\G'}(\tilde\BOmega)
,\etag\charp$$
where $P_{\G'}$ is an invariant symmetric tensor on $\Lie(SO(3))$.
This follows because
$\tilde\lambda(z,[u_i,u_j])$ equals a combination of vertical forms
(along the directions of the fiber $S(TM_z{}^{\uV'}/TM_z)$ being
integrated over) and the pull back of $\tilde\BOmega$ and because
this combiniation is invariant
under the $SO(TM_z)$ action on the $u_i$'s and on $\tilde\BOmega$.
Since $\tilde\lambda$ is universal, $P_{\G'}$ only depends on
$\G'$.

\bigskip

Now we come to the heart of the proof.
Up to now, our calculation of the anomaly would apply, with a little
modification, to calculating gauge fixing anomalies in a wide class of
theories; although the particular form for $A_{mfld}$ and $A_{gp}$ would
be different.
Now we will use those particular forms to prove
that $\bar A_{V',I'}$ vanishes unless $E=0$.
To begin, $\bar A_{V',I'}$ must have non-negative degree as a differential
form,
so
$E\le 4$.  Next, since $\tilde\BOmega$ has degree 2, \charp\ implies
that $E$ must be even if $\bar A_{V',I'}$ is to be nonzero.
This leaves only $E=2$ or $4$.  Finally, those case are handled
because $\Lambda^E(\Lie(G_{ss})^\dual)^{invt}$ is isomorphic to
the cohomology group $H^E(G_{ss};\IR)$.  By semisimplicity, the
latter group is trivial for $E=2$ or $4$.

\bigskip

When $E=0$, the terms involving $J$ in the expression
\expandl{.3}\ for $\tilde L_{reg,V''}$ do not contribute to
$\bar C_{V'',I''}$.  $\bar C_{V'',I''}$ is also independent of $z$.
In fact, changing labeling set
from $\{V'+1,...,V\}$ to $\{1, ...,V''\}$ on the right side
of \defC, we obtain the definition of one of the original perturbative
invariants,
$$\bar C_{V',I'} = I_{l''} \qquad\hbox{for $E=0$ and $l''=I''-V''+1$}
.\etag\last$$

Putting all of the above together, we have
\eqna\afive
$$\eqalignno{
  d_{\Met} I_l =& \sum_{l'=2\atop l''=l-l'}^l  A_{l'}\; I_{l''} &\afive{.1}
\cr
    A_{l'}  \equiv & \int_M \bar A_{2(l'-1),3(l'-1)}
       = -c_{V',I'}\sum_{\LG'} A_{gp}(\G')\int_M P_{\G'}(\tilde\BOmega),
								&\afive{.2}
}$$
where the sum is over labeled connected trivalent graphs with $l'$ loops.

In \afive{}, $P_{\G'}$ is an invariant tensor on $\Lie(SO(3))$ of degree
$2$.  This implies it must be a multiple of the
inner product.  So
$$ P_{\G'}(\tilde\BOmega)
 = {\alpha({\G'})\over 8\pi} <\tilde\BOmega,\tilde\BOmega>
\etag\last$$
for $\alpha({\G'})$ a constant which depends only on $\G'$.
But, with any choice of framing, the variation of the Chern-Simons action
of the metric connection is given by
$$ d_{\Met} \cs_{grav}(g,s)=
     {1\over 8\pi}\int_M 2 <\BOmega, \delta\Gamma>
    ={1\over 8\pi}\int_M <\tilde\BOmega,\tilde\BOmega>
.\etag\vcs$$
To obtain the last equality in \vcs,
recall (\eto{}) that
$$\tilde\Omega=\Omega+\delta\Gamma -\half\nabla (g^{-1}\delta g)
	       -\quarter (g^{-1}\delta g)\wedge(g^{-1}\delta g)
.\etag\last$$
The term involving
$\nabla\delta g$ vanishes by integration by parts and the Bianchi identity.

Putting the results of the last paragraph into \afive{.2}\ yields
$$\eqalign{
A_{l'} =& \beta_{l'}\; d_{\Met} \cs_{grav}(g,s)             \cr
\beta_{l'} =& -c_{V',I'}\;\sum_{\LG'} A_{gp}(\G')\; \alpha({\G'})
.}\etag\anok$$
$\beta_{l'}$ depends only on $l'$ and on the metric on $\Lie(G)$
(or even just its restriction to $\Lie(G_{ss})$).

The desired result
$$ d_{\Met} I_{l'}^{conn} = A_{l'} = \beta_{l'} d_{\Met}\cs_{grav}(g,s)
$$
follows from \afive{.1}, \anok, and the standard relation
$$ 1+\sum_{l=2}^\infty I_l k^{1-l}
  = exp\left({\sum_{l'=2}^\infty I_{l'}^{conn} k^{1-l'}}\right)
  \in \IR[[k^{-1}]]
\etag\conndisc$$
between sums over all graphs and connected graphs.

\bigskip

The only thing left now to complete the proof is to show
that $\beta_l$ vanishes for $l$ odd.
It suffices to show that the right hand side of \Ais{.2}\ vanishes
when $l'$ is odd.
This follows by looking at the involution $[\vec u]\rightarrow [-\vec u]$
of the integration region $S(TM_z{}^{\uV'}/TM_z)|_z$ in \Ais{.2}.
This involution is orientation reversing.
Also $\tilde\lambda$ is antisymmetric under the involution, as follows from
its explicit description (or the fact that $\tilde L$ is antisymmetric
under the involution of $M\times M$ exchanging the two copies of $M$).
Hence the integrand in \Ais{.2}\ is multiplied by
$(-1)^{I'}= -(-1)^{l'}$.  For $l'$ odd the integrand is invariant whereas
the orientation is not, so the integral in \Ais{.2}\ vanishes.

\bigskip

\noindent{\it Remark: }
Another way at arriving at the final result
for the sum $I_l^{conn}$ over connected graphs $\G$,
without using \conndisc, is to observe that all of our calculations
above for $d_{\Met} I_l$ apply to $d_{\Met} I_l^{conn}$ if we
omit terms coming from disconnected graphs $\G$.
To describe this, we need to use the graphical interpretation
of the sum \defC\ defining $\bar C_{V'',I''}$.
We will not elaborate on this now
except to say that $\bar C_{V'',I''}$ is given by a sum
over labeled graphs $\G''$ with $I''$ edges and with vertices labeled
from the set $\{z\}\union\uV''$.
Since the graphs $\G'$
summed over to yield $\bar A_{V',I'}$ are connected anyway,
the restriction that $\G$ be connected means that
the graphs $\G''$ must also be connected.
When, $E=0$, the vertex labeled by $z$ is always disconnected from
the rest of $\G''$, which must therefore be empty.  This means
that $V''$ equals zero.  So the only terms that contribute to the
anomaly are when $l''=0$ and $l'=l$.

\section\Appendix{Appendix}{Graded Tensor Product and Push-Forward Integrals}
\xdef\secsyme{A.}

Let $A^*$ and $B^*$ be graded algebras over $\IR$ with unit.
The graded tensor product $A^*\hat\otimes B^*$ is the tensor product
of the underlying vector spaces of $A^*$ and $B^*$ equipped with
the multiplication law given by
$$(a_1\otimes b_1)(a_2\otimes b_2)
   = (-1)^{|b_1||a_2|} (a_1 a_2)\otimes (b_1 b_2)
\etag\last$$
for $b_1\in B^{|b_1|}$ and $a_2\in A^{|a_2|}$ of pure degree, and
defined for all $b_1$, $a_2$ by linearity.
There is a natural graded algebra isomorphism
$A^*\hat\otimes B^*\cong B^*\hat\otimes A^*$ taking
$a\otimes b$ to $(-1)^{|a||b|} b\otimes a$ for $a$, $b$ of pure degree.
We write $a\otimes 1$ as $a$ and $1\otimes b$ as $b$, and multiplication
with or without a wedge product symbol,
e.g. $(a_1\otimes b_1)(a_2\otimes b_2) = a_1 b_1 a_2 b_2
 = a_1\wedge b_1\wedge a_2\wedge b_2$.
If $A^*=\Lambda^*(V)$ and $B^*=\Lambda^*(W)$ then there is a natural
graded algebra isomorphisms
$A^*\hat\otimes B^* \cong \Lambda^*(V\oplus W)$.

Suppose ${\bf B}^*\rightarrow Y$ is a bundle of graded algebras over
an oriented manifold $Y$.
We define multiplication of forms in $\Omega^*(Y, {\bf B}^*)$
by identifying this space with the graded algebra
$\Gamma(Y,\Lambda^*(T^*Y)\hat\otimes {\bf B}^*)$.

When ${\bf B}^*=Y\times B^*$ is a trivial bundle,
the algebras $\Omega^*(Y)\hat\otimes B^*$, $\Omega^*(Y,B^*)$, and
$\Omega^*(Y,{\bf B}^*)$ are by definition all equal.
Our notion of integration over $Y$ of such forms is defined by
$$ \int_Y \omega\wedge b = (\int_Y\omega) b \in B^*
\qquad\hbox{for $\omega\in\Omega^*(Y)$ and $b\in B^*$.}
\etag\last$$
and linearity.

Now suppose $X$, $Y$ are manifolds and ${\bf C^*}\rightarrow X$ is a
bundle of graded algebras.  Let $\pi_X:X\times Y\rightarrow X$ be the
projection map.  We may identify
$\Omega^*(X\times Y,\pi_Y^*({\bf C^*}))$ with a graded and completed
tensor product
$\Omega^*(X,{\bf C^*})\hat\otimes\Omega^*(Y)$.
Identifying $\Omega^*(X,{\bf C^*})$ with the algebra $B^*$ in the last
paragraph gives a notion of integration over $Y$ of forms in
$$\Omega^*(X\times Y,\pi_Y^*({\bf C^*}))
 \cong B^*\hat\otimes \Omega^*(Y)\cong \Omega^*(Y)\hat\otimes B^*
.\etag\last$$
Given $\rho\in\Omega^*(X\times Y,\pi_Y^*({\bf C^*}))$, let
$\psi=\int_Y\rho\in\Omega^*(X,{\bf C^*})$.  Then we write
$$\psi(x) = \int_{y\in Y} \rho(x,y)
  \in \Lambda^*(T^*X_x)\hat\otimes{\bf C}_x^*\quad\hbox{for $x\in X$.}
\etag\last$$
If $\rho(x,y)=\omega(x)\wedge\eta(y)$, then
$$\psi(x) =\int_{y\in Y} (\omega(x)\wedge\eta(y))
 = (-1)^{|\omega||\eta|}\int_{y\in Y} (\eta(y)\wedge\omega(x))
 = (-1)^{|\omega||\eta|}\left[\int_{y\in Y}\eta(y)\right] \omega(x)
.\etag\last$$
Note that this sign convention implies that
$$ D_X\int_Y \rho = (-1)^{\dim(Y)} \int_Y D_X\rho
,\etag\last$$
where $D_X$ is covariant exterior derivative operator in the $X$ directions
associated to a connection on ${\bf C^*}$.

\listrefs
\bye